\documentclass{aa}
\usepackage{psfig} 

\def\pks{PKS~0558-504}
\def\ergs{erg\,sec$^{-1}$}
\def\flux{erg\,cm$^{-2}$\,sec$^{-1}$}

\newcommand{\etal}{et al.}

\newcommand{\xmm}{{\it XMM-Newton}}
\newcommand{\rxte}{{\it RXTE}}

\newcommand{\ltsima} {$\; \buildrel < \over \sim \;$}
\newcommand{\gtsima} {$\; \buildrel > \over \sim \;$}
\newcommand{\lta} {\lower.5ex\hbox{\ltsima}}
\newcommand{\gta} {\lower.5ex\hbox{\gtsima}}

\def\approxlt{\mathrel{\hbox{\rlap{\lower.55ex \hbox {$\sim$}}
        \kern-.3em \raise.4ex \hbox{$<$}}}}
\def\approxgt{\mathrel{\hbox{\rlap{\lower.55ex \hbox {$\sim$}}
        \kern-.3em \raise.4ex \hbox{$>$}}}}

\begin{document}

\title{XMM$-$Newton long-look observation of the narrow line Seyfert~1 
       galaxy PKS~0558$-$504; I: Spectral analysis}  

\author{I.E. Papadakis\inst{1,2} \and W. Brinkmann\inst{3}  
  \and M. Gliozzi\inst{4} \and C. Raeth\inst{3} \and F. Nicastro\inst{2,5,6} 
  \and M. L. Conciatore\inst{6}}

\offprints{I.E. Papadakis;  e-mail: jhep@physics.uoc.gr}

\institute{Physics Department, University of Crete, P.O. Box 2208,
           GR--710 03 Heraklion, Crete, Greece
\and IESL, Foundation for Research and Technology, 711 10, Heraklion, Crete,
     Greece
\and Max--Planck--Institut f\"ur extraterrestrische Physik,
     Giessenbachstrasse, D-85740 Garching, Germany
\and George Mason University,
     Department of Physics and Astronomy, MS 3F3, 4400 University Dr.,
      Fairfax, VA 22030, USA
\and Osservatorio Astronomico di Roma, INAF, Italy 
\and Harvard-Smithsonian Center for Astrophysics, 60 Garden Street, 
     Cambridge, MA 02138, USA
}
 
\date{Received: ? / accepted: ? }

\abstract 
{\pks\ has been observed repeatedly by \xmm\ as a  calibration and performance
verification (PV) target.In this work, we present results from the spectral
analysis of a long XMM-Newton observation of the radio loud Narrow Line Seyfert
1 galaxy  PKS~0558-504.}
{To study the soft excess component in this object, the spectral variations it
exhibits in both the hard and soft X--ray bands, and their correlation.} {We
used mainly the PN data, and we fitted various spectral models to the time
average spectra of the individual orbits as well as the spectra from data
segments of shorter duration. We also used the RGS data to search for signs of a
warm absorber in the source.} 
{The source is highly variable, on all sampled time scales. We did not observe
any absorption features in either the soft or hard band. We found weak evidence
for the presence of an iron line at $\sim 6.8$ keV, which is indicative  of 
emission from highly ionized iron. The 2--10 keV band spectrum of the source is
well fitted by a simple power law model, whose slope steepens with increasing
flux, similar to what is observed in other Seyferts as well. The soft excess is
variable both in flux and shape, and it can be well described by a
low-temperature Comptonisation model, whose slope flattens with increasing flux.
Finally, the soft excess flux variations are moderately correlated with the 
hard band variations, and we found weak evidence that they are leading them by
$\sim 20$ ksec.}
{Our results rule out a jet origin for the bulk of the X--ray emission in this
object. We found no signals of a warm absorber. The observed hard band spectral
variations suggest intrinsic continuum slope variations, caused by changes in
the ``heating/cooling" ratio of the hot corona. The low-temperature 
Comptonising medium, responsible for the soft excess emission, could be a hot
layer in the inner disc of the source, which appears due to the fact that the
source is accreting at a super-Eddington rate.  The soft excess flux and
spectral variations could be caused by random variations of the accretion rate.
}

 
\titlerunning{X-ray variability of \pks}
\authorrunning{Papadakis~\etal}
\maketitle

\section{Introduction}  

Narrow Line Seyfert~1 galaxies (NLS1) are optically identified by their
emission  line properties: the ratio of [O III]/H$\beta$ is less than 3 and the
FWHM H$\beta$ less than 2000${~\rm km~s^{-1}}$ (Osterbrock \& Pogge 1985,
Goodrich 1989). Their optical spectra are further characterised by the presence
of strong permitted  Fe II, Ca II,  and O I $\lambda$ 8446 ionisation lines
(Persson 1988). NLS1 have been rarely found to be radio-loud (Ulvestad,
Antonucci, \& Goodrich 1995, Siebert et al. 1999, Grupe et al. 1999). Recent
dedicated studies, in particular using the Sloan digital sky survey,
demonstrated  that this is a consequence of the low radio-loud fraction of the
NLS1 population ($\sim$ 7\%, Komossa \etal\ 2006). The same authors found that
radio-loud NLS1 galaxies  resemble GHz-peaked/compact steep spectrum sources,
and Yuan \etal\ (2008) suggested similarities with high-energy peaked BL Lac
Objects.  

PKS 0558-504 ($z=0.1372, m_{\rm B}=14.97$) is one of the best studied
radio loud  NLS1 galaxies  ($R_{\rm L}=f_{\rm 5 GHz}/f_{\rm B}\simeq 27$;
Siebert et al. 1999).  A peculiar property displayed by the source is its
unusually  high X-ray to radio luminosity ratio (Brinkmann, Yuan, \& Siebert
1997). It was optically identified  on the basis of its X-ray position from
HEAO-1  (Remillard et al. 1986). It is an X--ray bright source, with  a flux of
$\sim 1.5-2\times 10^{-11}$ \flux\ in the 2--10 keV band. Its 2--10 keV spectrum
is rather steep ($\Gamma\sim 2.2;$ Gliozzi, Papadakis, \& Brinkmann 2007). Its
spectral slope  remains roughly constant on long time scales, despite the large
amplitude flux variations that the source exhibits (Gliozzi \etal\ 2007). \pks\
also exhibits strong flux variations on short time scales (Remillard et al.,
1991; Gliozzi et al., 2000; Gliozzi \etal, 2001; Wang \etal, 2001; Brinkmann
\etal, 2004).  

\pks\ has been observed repeatedly by \xmm\ as a  calibration and performance
verification (PV) target.  O'Brien et al. (2001) published the spectral analysis
of some preliminary  data of the commissioning/CalPV phase. The 0.2$-$2 keV
spectrum was dominated by a large soft X-ray excess,  which showed no evidence
for absorption or emission line features and  was fitted by a model consisting
of three black body components. An analysis of all PV phase and calibration
observations performed by \xmm\ up to  orbit 341, with various instrumental
settings, was presented by Brinkmann \etal\ (2004). They found that all spectra
could be well fitted by two Comptonisation components, one at moderate
temperatures of $k$T $\sim$ 4.5 keV and optical depths of $\tau \sim 2$, the
other at high temperatures ($k$T $\gta$ 50 keV) and low optical depths ($\tau
\lta 1.0$). The same \xmm\ data were recently used by Haba \etal\ (2008). The
spectra were  fitted with a multicolour disc black body model (Mitsuda \etal\
1984) plus a high energy power law. The results were interpreted in terms of a
super-critical  accretion flow with significant photon trapping in the inner
regions of the accretion disc.  

In September 2008, PKS 0558-504 was observed with \xmm\ for five consecutive
orbits. The main objective of this long observation was to study its X--ray flux
variability and in particular to investigate whether it exhibits any
quasi-periodic signals. The results from this study will be presented in a
companion paper (Papadakis \etal, in preparation). In this paper we present the
results from the  spectral analysis of the \xmm\ data. We used mainly the PN
data because of the higher sensitivity of the PN camera, although we sometimes
resorted to the RGS and MOS data as well, mainly for clarification purposes. We
start with the observational details in the next section.  Section 3 deals with
the spectral  analysis of the PN data, first in the hard band and then for the
total band. In the same section we also present the results from a preliminary
analysis of the RGS data. Temporal spectral variations and a time resolved
spectroscopic analysis are discussed in Sects. 4 and 5. We then present a more
detailed  analysis of the full band spectra in selected time intervals in Sect.
6.  Finally, in Sect. 7 we summarise our results and we present a short
discussion of their potential implications. 

%%%%%%%%%%%%%%%%%%%%%%%%%%%%%%%%%%%%%%%%%%%%
\section{Observations and data analysis}
%%%%%%%%%%%%%%%%%%%%%%%%%%%%%%%%%%%%%%%%%%%%

%%%%%%%%%%%%%%%%%%%%%%%%%%% FIG 1 %%%%%%%%%%%%%%%%%%%%%%%%%%%%%
\begin{figure*}[t]
\psfig{figure=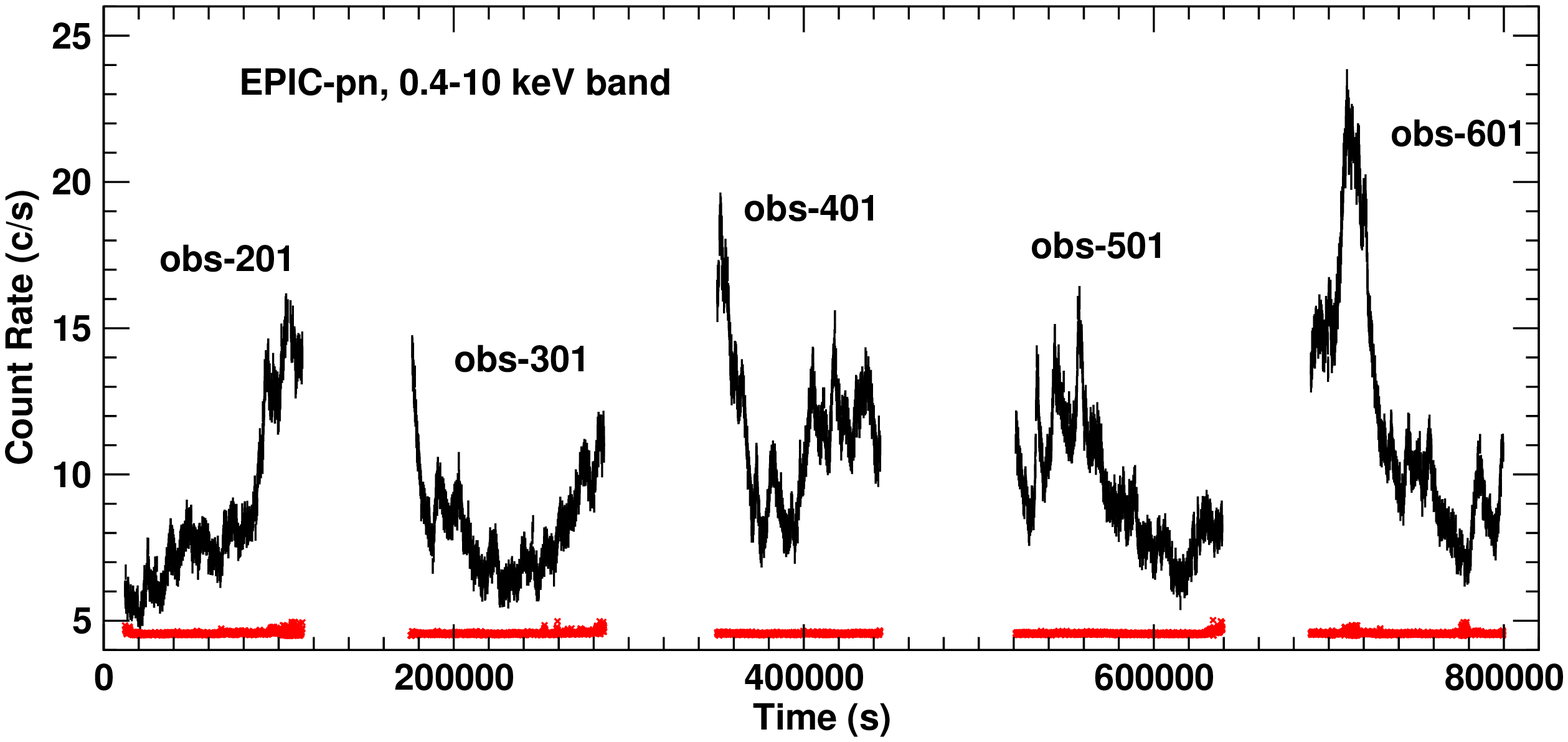,width=16.5cm,height=10.6cm,%
 bbllx=44pt,bblly=169pt,bburx=768pt,bbury=510pt,clip=}
 \caption {The PN 0.4$-$10 keV light curve of PKS 0558-504 for all five
observations with a binning of 100\,sec. The time counts from the beginning of
the first observation. The points plotted with crosses at the bottom of the plot
indicate the background count rate  in the same energy range.}             
\label{figure:tot-lc} 
\end{figure*}

PKS 0558-504 was observed with all instruments on board XMM$-$Newton for five
consecutive orbits starting on 2008 September 9, 1:19 UT and ending on 2008
September 16, 12:02 UT. The observation identifiers range from 0555170201 (XMM
revolution 1602)  to 0555170601 (XMM revolution 1606). Below we will identify
the individual observations by the last 3 digits (i.e. `obs-201', `obs-301',
etc). The PN was operated in Small Window mode (SW) with a thin filter, the MOS1
in Small Window mode with a thin filter, and the MOS2 in TIMING mode also thin
filter. Both RGS were operated in spectroscopy mode.  

For the following analysis we will rely mainly on data from the PN camera
(Str\"uder et al. 2001) due to its superior statistical quality. The PN data
have been processed using {\small XMMSAS} version 8.0.  Source counts were
accumulated from a rectangular box of 27$\times$26 {\small RAW} pixels  (1
{\small RAW} pixel $\sim$ 4.1\arcsec) around  the  position of the source.
Background data were extracted from a similar, source free region on the chip. 
We selected single and double events ({\small PATTERN}$\leq$4, {\small
FLAG}$=$0; for details of the instruments see Ehle \etal\ 2008). With an
average  count rate of about 10 cts~sec$^{-1}$ photon pile-up is negligible for
the PN detector, as verified using  the {\small XMMSAS} task $epatplot$.  The
data were screened for high background in both the soft and hard bands. After
rejection of time intervals affected by high background (usually at the end of
the individual orbits) and flaring activity, the total useful observation time
was 540 ksec. For the spectral fits we employed for the PN the canned responses
from the MPE database. 

As for the RGS data, for each observation  we used the tool $rgsproc$ to extract
calibrated first order spectra and responses for both RGS cameras.  The RGS
spectra are affected by numerous instrumental features due to hot pixels and/or
bad columns  in the read-out chips. To reduce the number of narrow spectral
intervals blocked by such features, and their width,  we modified the default
$rgsproc$ filtering parameters to exclude photons only from hot pixels and not 
from their neighbouring pixels.  The RGS data can be affected by high particle
background periods during parts of the \xmm\ orbits, mostly caused by solar
activity. The high energy band is the most affected by background flares, and 
high energy RGS photons are dispersed over the CCD-9 chip. We therefore
extracted background light curves of the  CCD-9 chip and selected as
good-time-intervals of the processed observations only those during which the
background count rate deviated by $\le +2\sigma$ from the average background
count rate of each observation.  To improve and optimise the S/N of the RGS
spectra, we then used the tool $rgscombine$ to co-add the five RGS1 and RGS2
spectra. The final (after cleaning for high background time intervals) exposures
of the RGS1 and RGS2 spectra are 480 ksec and 466 ksec, respectively. 

The 0.4$-$10\,keV, PN light curve is shown in  Fig.~\ref{figure:tot-lc}. The
data,  plotted in 100 sec time bins, represent the actual count rate on the
chip, i.e. they are not corrected for the live time of 0.71 of the SW mode. The
source exhibits large amplitude flux variations on all sampled time scales. On
long time scales,  the source undergoes large intensity variations by a factor
of $\leq$ 5 (from $\sim 5$ counts/sec at the start of obs-201 to $\sim$ 24
counts/sec at the beginning of obs-601).  On shorter time scales of the order of
a few hours we still observe substantial  and continuous  variability  with an
amplitude of the order of $\sim 2$. A detailed analysis of the observed flux
variations will be presented in a forthcoming paper. 

\section{Spectral analysis of the PN data}

Previous studies of \pks\ have confirmed the presence of spectral components
which are typical of the X--ray spectra of NLS1 galaxies: i) a strong soft
excess below $\sim$ 2 keV, which appears over the extrapolation of a higher
energy power-law-like spectral component with a slope of $\sim 2-2.4$  (Gliozzi
et al., 2000; O'Brien et al., 2001; Brinkmann \etal, 2004), and ii) a smooth
spectrum, with no obvious absorption and/or emission spectral features either at
higher energies or in the soft band.  Given the presence of the soft component,
we first investigated the spectral shape of the source spectrum  at energies
above 2 keV. We then proceeded to study the full band energy spectrum. 

For the spectral analysis, source counts were grouped with a minimum of 100
counts per energy bin. Spectral fits were performed with the {\tt XSPEC}
v.12.5.0 software package. The errors on the best-fitting model parameters
represent the 90\% confidence limits for one interesting parameter (unless
otherwise stated). The energies of emission or absorption features are given in
the rest frame of the source. We considered a model to provide an acceptable fit
to the data if the quality of the fit is better than the 5\% confidence level,
and we accepted that the addition of a model component is significant if the
quality of the fit is improved at more than the 95\% confidence level.  For the
Galactic absorption we used the value of  N$_{\rm H}=3.4\times 10^{20}$
cm$^{-2}$ (Kalberla \etal\ 2005).

\subsection{High energy power law fits}
 
\begin{figure}
\psfig{figure=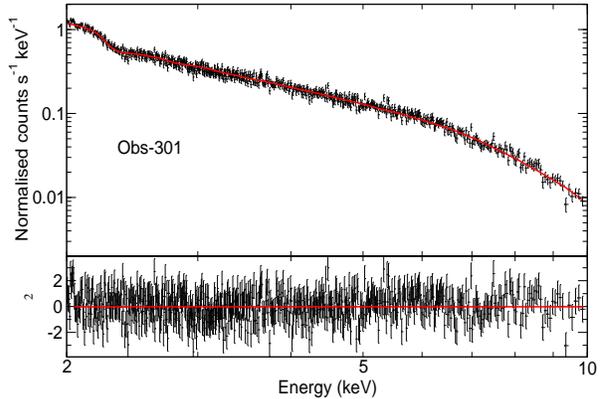,width=8cm,height=5.5cm,%
 bbllx=0pt,bblly=40pt,bburx=770pt,bbury=530pt,clip=}
\caption{Top panel: Power law model fit to the total spectrum of obs-301 in the
2--10 keV band. Lower panel: The best-fit residuals plot (residuals are  in
terms of $\sigma$'s with error bars of size one).} 
\label{figure:orb301fit}
\end{figure}

We first considered the average spectrum from each individual \xmm\
observation.  A simple power law model (PL) fitted well the  2$-$10 keV spectrum
of all observations, except for obs-301.  That spectrum is rather noisy. 
Figure~\ref{figure:orb301fit} shows the best power law model fit for this
observation and the corresponding residuals. A broad, low-amplitude, line-like
feature does appear close to $\sim 6$ keV (observers frame), but the addition of
a Gaussian component did not improve the quality of the fit significantly (see
the discussion in the next section). Possible  absorption features also appear
around $\sim 7$ keV. The addition of an {\tt EDGE} component (in {\tt XSPEC}
terminology) to the model improved the fit (we got a reduction of $\Delta\chi^2$
of 24.8 for the addition of 2 degrees of freedom, dof), however, the best-fit
edge energy turned out to be $\sim 2.9$ keV. It is difficult to spot this
feature in Fig.~\ref{figure:orb301fit}, and we did not detect a similar feature
in any of the average spectra of the other four observations. This fact casts
doubt on the possibility that such an absorption edge is an intrinsic feature in
the spectrum of the source.

%%%%%%%%%%%%%%%%%%%%%%%%%%% FIG 2 %%%%%%%%%%%%%%%%%%%%%%%%%%%%%%%%%
   
The PL best-fit results are listed in Table \ref{table:fithard}. All spectra
show a slope of $\Gamma \sim 2.13-2.15 $. We did not observe  significant
indications of any absorption features in the high energy band spectra. We did
not observe any spectral  curvature in the hard energy band either.
Nevertheless, a Comptonisation model ({\tt compTT} in {\tt XSPEC}, Titarchuk
1994) provided a nearly identical fit to the data. The main model parameters are
the temperature, T$_0$, of the soft photon input spectrum (which is assumed to
be the Wien part of a black body), the electron  plasma temperature, $k$T, and
optical depth, $\tau$. Their best fit values  (T$_0 \sim 65$ eV, $k$T $\sim$ 50
keV, and $\tau \sim 0.5$)  are subject to large uncertainties due to their
inherent degeneracy (in the absence of a the detection of the high energy
cut-off) but, in any case, they are similar to those found for broad line
Seyfert~1 galaxies (e.g. Petrucci et al. 2001). 

The average $2-10$ keV flux of the source during the \xmm\ observations is $\sim
1.1\times 10^{-11}$ \flux. This implies that \pks~ was at a rather low flux
state, when compared to its average flux level of $\sim 1.8\times 10^{-11}$
\flux, as determined from its long term \rxte\ monitoring observations (Gliozzi
\etal\ 2007). At a distance of 624 Mpc \footnote{This luminosity distance
estimate is taken  from NASA/NED. It is calculated assuming H$_{\rm 0}=73$
km/sec/Mpc, $\Omega_{\rm matter}=0.27$, and $\Omega_{\rm vacuum}=0.73$.} this
flux corresponds to an average luminosity of $\sim 5.6 \times 10^{44}$ \ergs\
during the \xmm\ observations.  

\begin{table}[]
\small
\tabcolsep1ex
\caption{\label{table:fithard} Results from the PL model fits to the 2$-$10 keV 
average spectra of the 5 observations. }
\begin{tabular}{lccc}
\noalign{\smallskip} \hline \noalign{\smallskip}
\multicolumn{1}{c}{Obs} &
\multicolumn{1}{c}{$\Gamma$} &
\multicolumn{1}{c}{2--10 keV Flux}$^\dagger$& 
\multicolumn{1}{c} {$\chi^{2}_{\rm red}/$dof} \\

\noalign{\smallskip} \hline \noalign{\smallskip}
201 & 2.131$\pm$0.015&9.53$\pm$0.05& 1.036/591 \\
\noalign{\smallskip}
301 & 2.142$\pm$0.014&9.07$\pm$0.05& 1.119/617 \\
\noalign{\smallskip}
401 & 2.154$\pm$0.011&13.0$\pm$0.01& 1.053/820 \\
\noalign{\smallskip}
501 & 2.138$\pm$0.012&11.2$\pm$0.01& 1.063/763 \\
\noalign{\smallskip}
601 & 2.150$\pm$0.011&13.9$\pm$0.01& 1.066/797  \\
\noalign{\smallskip}\hline
\end{tabular}
\medskip
  
$^{\dagger}$in units of $10^{-12}$ \flux 
\end{table}

\begin{table}[]
\small
\tabcolsep1ex
\caption{\label{table:fitline} The iron-line best fit parameter values.}
\begin{tabular}{lcrc}
\noalign{\smallskip} \hline \noalign{\smallskip}
\multicolumn{1}{c}{Obs} &
\multicolumn{1}{c}{E$_{\rm line}$ (keV)} &
\multicolumn{1}{c}{$\Delta\chi^2$} &
\multicolumn{1}{c}{EW (eV)} \\

\noalign{\smallskip} \hline \noalign{\smallskip}
201 &  6.8 & 1.1~(175) & 8~($<33$) \\
\noalign{\smallskip}
301 &  $6.69\pm 0.07$ & 5.7~(188) & 18~($<42$) \\
\noalign{\smallskip}
401 &  $6.95\pm 0.05$ & 10.8~(276) & 19~($<38$) \\
\noalign{\smallskip}
501 &  $6.90\pm 0.12$ & 4.1~(251) & 11~($<32$) \\
\noalign{\smallskip}
601 &  $6.83\pm 0.05$ & 5.9~(264) & 13~($<31$) \\
\noalign{\smallskip}
Total &  $6.82\pm 0.10$ & 22\ ~(697) & 23$^{+25}_{-22}$ \\
\noalign{\smallskip}\hline
\end{tabular}
\medskip
\end{table}
  
\subsection{The iron line}

%%%%%%%%%%%%%%%%%%%%%%%%%%%%%%%% FIG 3 %%%%%%%%%%%%%%%%%%%%%%%%%%%%%%%%%
\begin{figure}
\psfig{figure=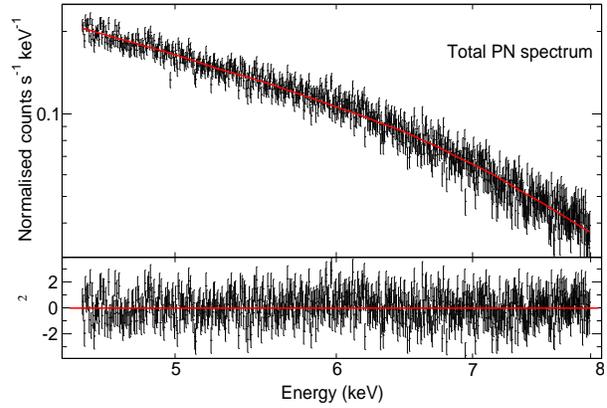,width=8cm,height=5.5cm,%
bbllx=15pt,bblly=37pt,bburx=758pt,bbury=525pt,clip=}
\caption{Top panel: The best power law model fit to the total PN spectrum in
the 4.5--8 keV band. Lower panel: The best-fit residuals plot.} 
\label{figure:feline}
\end{figure}

%%%%%%%%%%%%%%%%%%%%%%%%%%%%%%%% FIG 4 %%%%%%%%%%%%%%%%%%%%%%%%%%%%%%%%%
\begin{figure*}[t]
\psfig{figure=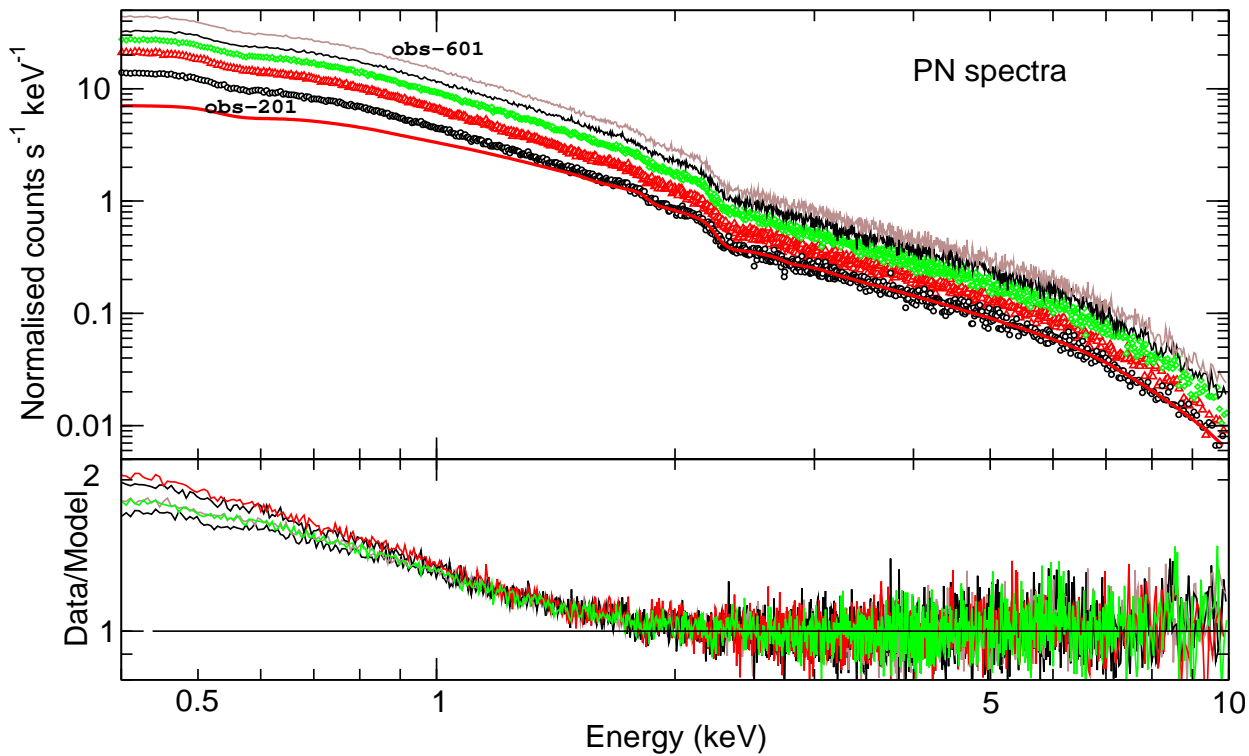,height=10cm,width=16.5cm,%
bbllx=64pt,bblly=125pt,bburx=727pt,bbury=535pt,clip=} \caption{Top panel: The
0.4--10 keV PN spectra of all observations.  Black, open circles indicate the
obs-201 spectrum; red, open triangles indicate the obs-301 spectrum, and green,
open diamonds indicate the obs-401 spectrum. The obs-501 and obs-601 spectra are
plotted with a continuous black and brown line.  Some of the spectra are scaled
for demonstration reasons (see text for details). The solid line indicates the 
best PL fit to the obs-201, 2--10 keV spectrum. Lower panel: The ``Data/the best
hard band PL model fit" plots for the same spectra (plotted using the same
colour code as in the top panel).}
\label{figure:allspec}
\end{figure*}
 
The previous \xmm\ data of \pks\ yielded low upper limits on the equivalent
width of an iron line, either from neutral or ionised material (O'Brien \etal\
2001; Brinkmann \etal\ 2004). The present data are consistent with the results
from these studies: the best PL fit residuals to the individual spectra of the
five observations do not show clear evidence for the presence of an iron line.

To investigate this issue in a more quantitative way, we re-fitted the five
average spectra of each observation with the PL model in the 4.5--8 keV band. We
fixed the best-fit spectral slope and then added a narrow Gaussian component to
the model (the line width was kept fixed at a value of $\sigma=10^{-5}$ eV). The
best-fit results are listed in Table \ref{table:fitline}. The third column in
this table lists the improvement in the best-fit $\chi^{2}$ (i.e.
$\Delta\chi^2$) when we added the Gaussian component to the model. The numbers
in the parentheses indicate the degrees of freedom in the case of the PL model;
the addition of the Gaussian component reduces this number by 2. The small
$\Delta\chi^2$ values  suggest that the addition of the Gaussian component is
not statistically significant (except perhaps in the case of obs-401). The
numbers in the parenthesis in the fourth column indicate the 90\% upper limits
on the equivalent width (EW) of the line. In fact, the best-fit E$_{\rm line}$
values turned out to be close to the starting values we chose. The rather low
90\% upper limits imply that if there is an iron line in the energy spectrum of
\pks, it is weak. 

For that reason, we combined the data from all orbits into a single spectrum 
and fitted it again with a power law model in the 4.5 to 8 keV energy range. 
Figure~\ref{figure:feline} shows the best power law model fit to the data,
together with the residuals plot. The reduced $\chi^2$ value of 767/697 dof
implies a null hypothesis probability of 3.4\%. The residuals indicate a
low-amplitude excess emission around $\sim 6$ keV.  As before, we added a
Gaussian component to the model and repeated the fit (keeping the best-fit
spectral slope fixed). In this case, we also let the line's width, $\sigma$,  be
a free parameter. The best-fit results are also listed in Table
\ref{table:fitline} (the best-fit $\sigma$ value was  $0.19^{+0.24}_{-0.09}$
keV). The line's best-fit energy is indicative of emission from Fe XXV. However,
the large error of the line's equivalent width measurement suggests that the
detection of the line itself is only at the 90\% significance level. We
therefore conclude that if there is an iron line in the spectrum of the source
it is very weak (EW$\sim 20$ eV), and an indication of emission from ionized
material.  

\subsection{The full band EPIC-PN spectrum}

An extrapolation of the hard power law to low energies shows a clear ``excess"
of flux below $\sim$ 2 keV. This is demonstrated in Fig.~\ref{figure:allspec}. 
In the top panel we show the spectra of all observations. For demonstration
reasons,  the obs-601 and obs-501 spectra are multiplied by 1.5, while the
obs-201 spectrum has been divided by 1.5. In this way, the spectra plotted from
bottom to top correspond to the data from obs-201 to obs-601, in increasing
order. The spectra have been fitted with a PL model in the 2--10 keV band (the
solid line in the same panel indicates the best PL model fit to the obs-201,
hard band  spectrum).  The bottom panel in Fig.~\ref{figure:allspec} shows the
``data over the  best hard band PL fit" ratio. It is difficult to spot the data
for the individual observations at energies higher than 2 keV in this panel, 
because the PL model fits all spectra  well at these energies.  As a result the 
data/model ratio is close to unity in all of them. However, at energies below
$\sim 2$ keV, a strong, featureless excess on top of the extrapolation of the
2$-$10 keV PL fit is clearly evident.  The soft excess is almost 1.5-2 times
``stronger" than the extrapolation of the hard band power law continuum.
Interestingly, the data plotted in the lower panel of Fig.~\ref{figure:allspec}
indicate that the amplitude of this excess emission is variable. 

Various models have been proposed for the ``soft excess" component in NLS1
galaxies. In \pks\ in particular,  O'Brien et al. (2001) fitted three black body
components to this excess, Brinkmann \etal\ (2004) used an extra Comptonisation
component, while Haba \etal\ (2008) used  a multicolour disc black body model.
Finally, recent claims about the connection  of strong radio loud NLS1 galaxies
with BL Lac objects (Yuan \etal\ 2008) raise the possibility that the X--ray
spectra in objects like \pks\ could be modelled by broken power laws.

\begin{figure}
\psfig{figure=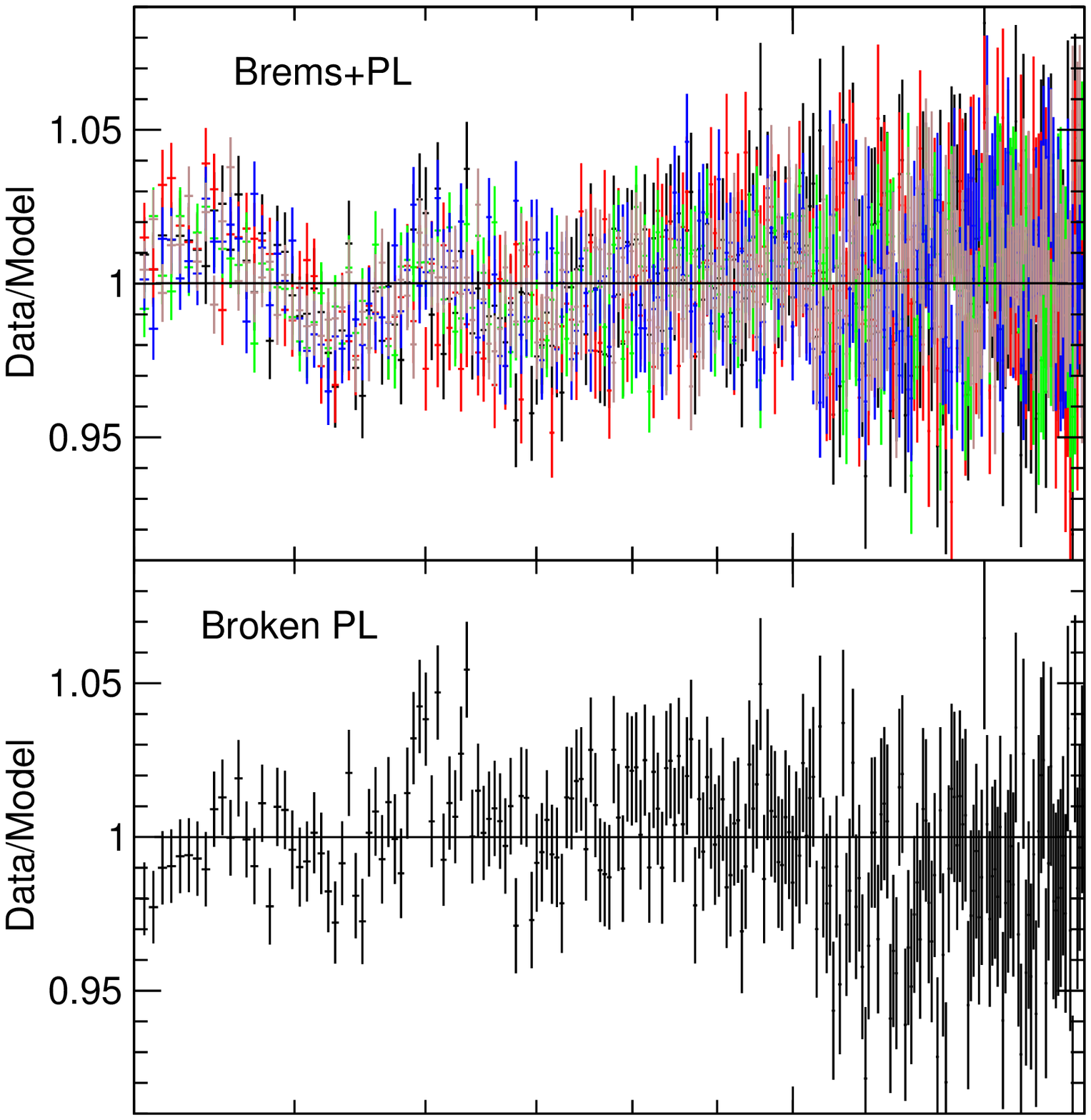,height=7.5truecm,width=8.5truecm,%
 bbllx=25pt,bblly=267pt,bburx=508pt,bbury=760pt,clip=}
\psfig{figure=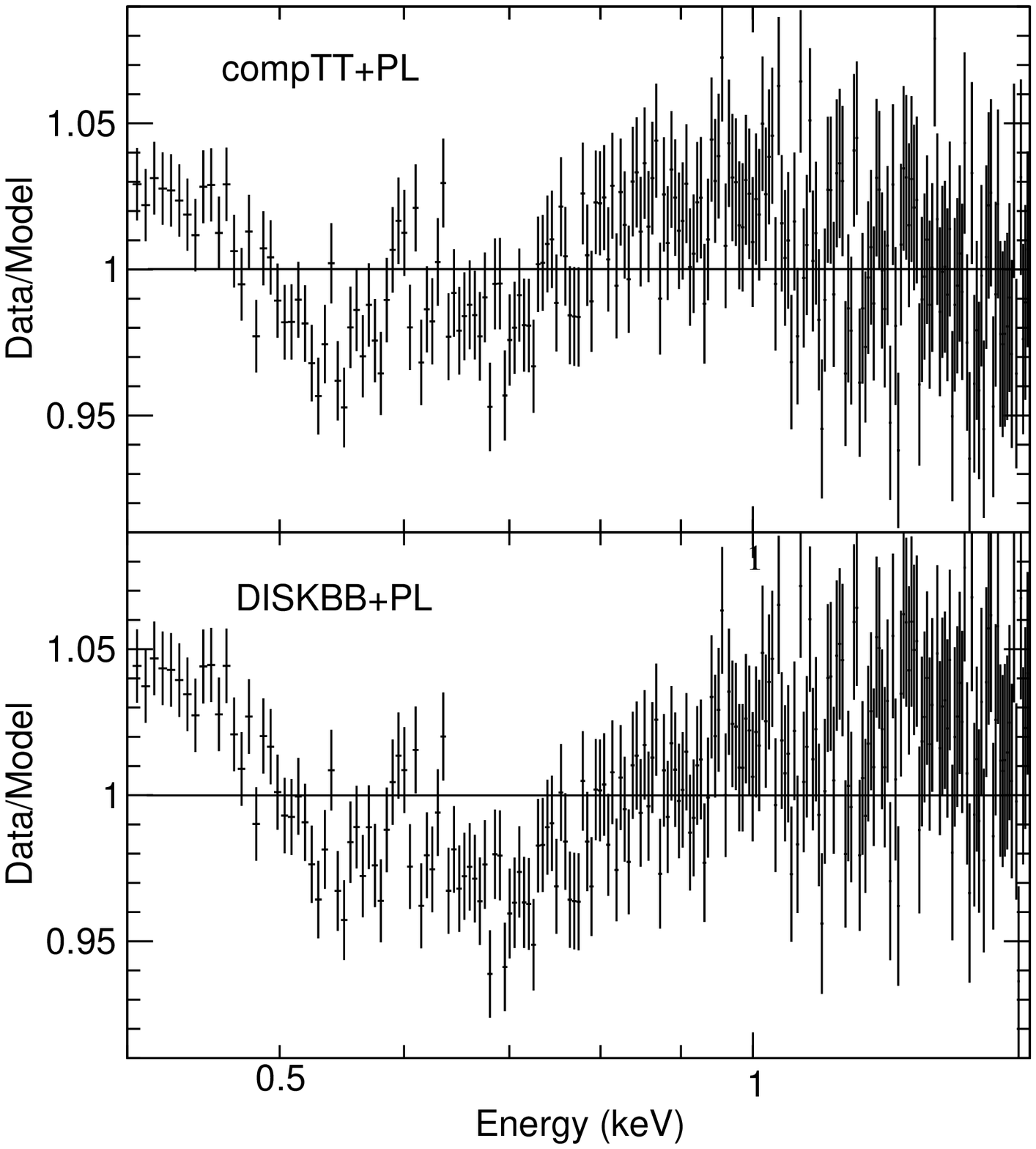,height=7.9truecm,width=8.5truecm,%
 bbllx=25pt,bblly=230pt,bburx=508pt,bbury=755pt,clip=}
\caption{ From top to bottom: Data-to-model ratio in the 0.4$-$2 keV band  for
the best fits of a Brems+PL, a broken PL, a {\tt compTT}+PL, and a {\tt
DISKBB}+PL model. In the top panel, we plot the data for the  the average
spectrum of each \xmm\ observation (black, red, green, blue and brown points
have been used for the obs-201, obs-301, obs-401, obs-501, and obs-601 spectra,
respectively). For clarity reasons, in the following panels, we plot the data
for the obs-201 spectrum only (the data-to-model ratios are very similar for the
other spectra as well).}
\label{figure:residua}
\end{figure}

To check which of the proposed models provides the best interpretation of the
source's X--ray spectrum, we fitted the spectra from the five individual
observations with various models. In Fig. \ref{figure:residua} we show the
data-to-model ratios for fits of four different models. The fits were performed
in the 0.4--10 keV band, by keeping the PL spectral slopes fixed at the values
listed in Table~\ref{table:fithard}. 

In the top panel of Fig.~\ref{figure:residua} we show the data-to-model ratio
when we used a bremsstrahlung component to account for the soft excess
(``PLBrems" hereafter, for  the combination of a ``hard power law plus a
bremsstrahlung component").  The residuals are similar in all five spectra (for
this reason it is difficult to spot the residuals for the individual spectra in
this panel). The best-fit resulted in a $\chi^2$ of 5796 for 5203 dof and is
formally not acceptable. The residuals show small amplitude absorption
structures near the O{\small I} edge and perhaps near $\sim 0.7$ keV as well
(observers frame). 

In the following panel we present the data-to-model ratio plot for the case  for
which  we fitted the full band spectrum with a broken PL model (``BKNPL",
hereafter). For clarity reasons,  we plot the data-to-model ratio only for the
obs-201 spectrum in this panel as well as the remaining panels of
Fig.~\ref{figure:residua}.  The model residuals are almost identical for the
other spectra as well. We found that the fit was slightly worse in this case
($\chi^2=6196/5203$ dof). The best-fit soft band slope and break energy were 
varying between 2.5 and 2.7, and 1.8--2 keV, respectively. 

In the bottom two panels, we show the data-to-model ratios when we used two
``physical" models, namely the {\tt compTT} and {\tt DISKBB} models in {\tt
XSPEC}, to account for the soft excess. In the case of the {\tt compTT}+PL model
($\chi^2=6863/5293$ dof), we first fitted the spectra by letting T$_{\rm 0}$
free to vary. The resulting best-fit values were similar for the five spectra,
and for this reason we repeated the fit by forcing T$_{\rm 0}$ to be the same in
all cases. The best fit T$_{\rm 0}$ value turned out to be $\sim 0.14$ keV,
while $k$T, and $\tau$ were varying between $\sim 2-4.4$ keV and $\sim 1.5-2.6$,
respectively. In the case of the {\tt DISKBB}+PL model  ($\chi^2=8024/5203$
dof), the best-fit temperature at the inner radius was $\sim 0.16-0.18$ keV.  We
also tried models like a combination of two power laws, or a black body plus a
hard band PL, but these models resulted in significantly worse fits. 

%%%%%%%%%%%%%%%%%%%%%%%%%%%%%%%% FIG 5 %%%%%%%%%%%%%%%%%%%%%%%%%%%%%%%%%
 
\begin{figure}
\psfig{figure=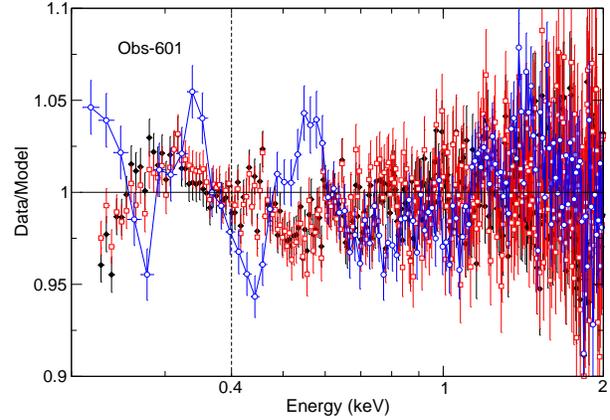,height=5.5truecm,width=8truecm,%
 bbllx=15pt,bblly=47pt,bburx=710pt,bbury=530pt,clip=}
\caption{``Data-to-PLBrems model" ratio in the 0.25--2 keV band, for the single
PN events  (open squares),  single plus double PN events (filled  circles), and
the MOS1 spectra (open circles) of obs-601.}
\label{figure:compare}
\end{figure}

The amplitude of the residuals becomes larger at energies below $0.4$ keV. Using
the whole energy band, i.e. $\ge$ 0.2 \,keV, yielded much larger best fit
$\chi^2_{red}$ values. This indicates either a failure of the models at the
lowest energies, or an inherent detector calibration uncertainty. To check the
possibility of calibration uncertainties, we additionally performed fits (using
the PLBrems model) with the single events (PATTERN = 0) of the PN and with the
MOS1 data. Open circles, open squares and the filled diamonds in
Fig.~\ref{figure:compare} indicate the 0.25--10 keV best-fit data-to-model
residuals in the case of obs-601 for the MOS1, the single, and the 
single+double PN events  spectra, respectively (we obtained similar results for
the PN and MOS1 spectra of the other observations as well). The residuals for
the  single and single+double PN spectra are quite similar. On the other hand,
there are differences between the  PN and the MOS1 residuals below $\sim 0.6$
keV. At energies $< 0.4$ keV, the MOS1 and PN residuals are quite different.
Their max-to-min difference  is of the order of $\sim 10$\%, a result which 
indicates that there exist unresolved detector calibration uncertainties at
these low energies (it is for this reason that we considered the PN spectra at
energies above 0.4 keV).  

At higher energies, the difference between the MOS1 and the PN spectra is less
than $\sim 5$\% except for the 0.5--0.6 keV band (where we expect the
instrumental and the Galactic oxygen edges to appear). This result implies that
any ``dips" in the residual plots at these energies  (like the ones shown in
Fig.~\ref{figure:residua}) may be due to additional calibration uncertainties. 
Furthermore, the {\tt compTT}+PL and {\tt DISKBB}+PL residual plots in
Fig.~\ref{figure:residua} exhibit a further shallow flux deficit around $\sim
0.7$ keV, which may correspond to low amplitude absorption features intrinsic to
the source. This possibility can be investigated with the study of the RGS
spectrum, which we describe below.

%%%%%%%%%%%%%%%%%%%%%%%%%%%%%%%% FIG 6 %%%%%%%%%%%%%%%%%%%%%%%%%%%%%%%%%

\subsection{The RGS data}

We used the fitting package {\em Sherpa}, of the Chandra Interactive Analysis of
Observation ({\em Ciao}) software (vs. 4.1.2), to simultaneously fit the total
RGS1 and RGS2 spectra of \pks. Both the  RGS1 and RGS2 spectra are  grouped at a
resolution of 20 m\AA\ ($\sim 0.5$ eV at 0.5 keV), thus allowing for three bins
for a RGS resolution element.  For both the RGS1 and the RGS2 the photon count
rate of the background spectrum becomes comparable to the background-subtracted
source count rate at E$\lta 0.4$ keV ($\lambda\ge 30$ \AA), and E$\gta 2$ keV
($\lambda\le 6.2$ \AA).  Moreover, due to failures of two different read-out
detector chips early in the mission, both the RGS1 and the RGS2 lack response in
two different spectral intervals: $\sim 0.9-1.2$ keV (RGS1) and $\sim 0.5-0.7$
keV (RGS2). Finally, a visual inspection of the RGS2 spectrum and its effective
area revealed several regions where the modelling of effective area features
failed to properly reproduce the data. These were mostly concentrated at E$\lta
0.5$ keV and in two narrow regions at E$\sim 0.84$ keV and E$\sim 1.08$ keV.  We
therefore considered the following two RGS1 and RGS2 spectral intervals for
spectral fitting purpose:  $(0.4-0.9) + (1.2-2)]$ keV and $(0.69-0.82) +
(0.85-1.07) + (1.09-2)]$ keV, respectively.

Due to the EPIC-RGS cross calibration problems, the RGS data cannot be compared
directly to the PN best-fit models. However, this is not a serious problem, as
our main aim in this work is to identify any strong emission and/or absorption
features in the RGS spectra that should be included in the models we use to fit
the PN spectra.  To this end, all we need is to model the continuum in the RGS
spectrum as accurately as possible and search for any remaining residuals.

\begin{figure}
\psfig{figure=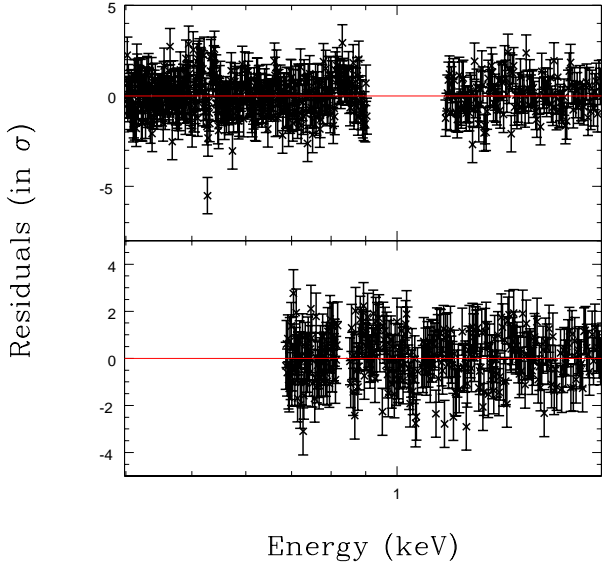,width=8cm,height=7.5cm,%
bbllx=18pt,bblly=170pt,bburx=550pt,bbury=675pt,clip=}
\caption{RGS1 (top panel) and RGS2 (bottom panel) residuals to a broken power
law fit attenuated by Galactic absorption. For clarity reasons,  the data
in this figure are binned  by a factor of 8 when  compared to the unbinned RGS
pipeline data. Consequently,  the spectra plotted in this figure are grouped at 
about 40 \AA per resolution element, i.e. at about 1.5 bins per RGS resolution
element.}
\label{figure:rgs}
\end{figure}

We first fitted a simple power law attenuated by neutral absorption, set to be
at least equal to the Galactic column along the line of sight. Although the fit
is statistically acceptable, $\chi^2=936.4/918$ dof, a visual inspection of the
broad band 0.4--2 keV residuals shows a flattening at energies higher than
$\sim  1.2$ keV, both in the RGS1 and RGS2. We therefore replaced the PL  with a
broken power law model. This fitted  the 0.4$-$2 keV continuum in the RGS
spectra very well ($\chi^2=874.4/914$ dof). The best-fit residuals are shown in
Fig.~\ref{figure:rgs}.

%%%%%%%%%%%%%%%%%%%%%%%%%%%%%%%% FIG 7 %%%%%%%%%%%%%%%%%%%%%%%%%%%%%%%%%

The residuals do not show any significant spectral features  identifiable with
either absorption or emission transitions within $\pm 5000$ km/sec from the
systemic redshift of PKS 0558-504, that should be considered in the PN spectral
fits. In particular there are no signatures of ionised material (i.e. the
so-called ``warm absorbers"; WA) which are observed in other Seyferts. We tried
to investigate this in a more quantitative way by adding {\em PHASE},  i.e. the
warm absorber model of  Krongold et al. (2003), to the broken power law model. 
We fixed the turbulence velocity of the absorber to 200 km s$^{-1}$ (unresolved
in the RGS), and left the ionisation parameter $U$ (defined as the ratio between
the photon volume density at the location of the absorbing cloud and the
electron density in the cloud) and the equivalent column density  of the
absorber, N$_H$, free to vary. The redshift of the absorber was fixed to the
systemic redshift  of PKS~0558-504, and its outflow/inflow velocity was allowed
to vary between -5000 km s$^{-1}$ and +5000 km s$^{-1}$. 

The best model fit was achieved for log(N$_H$)=20 (the lowest tabulated value in
the model) and log$(U)=1.37$. At these column densities and for this ionisation
parameter, the gas is already fully transparent in the RGS band. The best-fit
parameter values are not well determined, but the allowed parameter values (for
example log$(U)>2$ for log(N$_H$)=20--21, and log$(U)>2.3$ if log(N$_H$)=21--22)
are incompatible with the parameter space which would be normally occupied by a
WA (i.e. log(N$_H$)=20--22 and log$(U) <$1.1--2). We therefore conclude that the
presence of a warm absorber is not required by these data. A more detailed study
of the RGS spectrum of the source is presented in Nicastro et al. (submitted to
ApJ). 

%%%%%%%%%%%%%%%%%%%%%%%%%%%%%%%% FIG 8 %%%%%%%%%%%%%%%%%%%%%%%%%%%%%%%%%
\begin{figure}
\psfig{figure=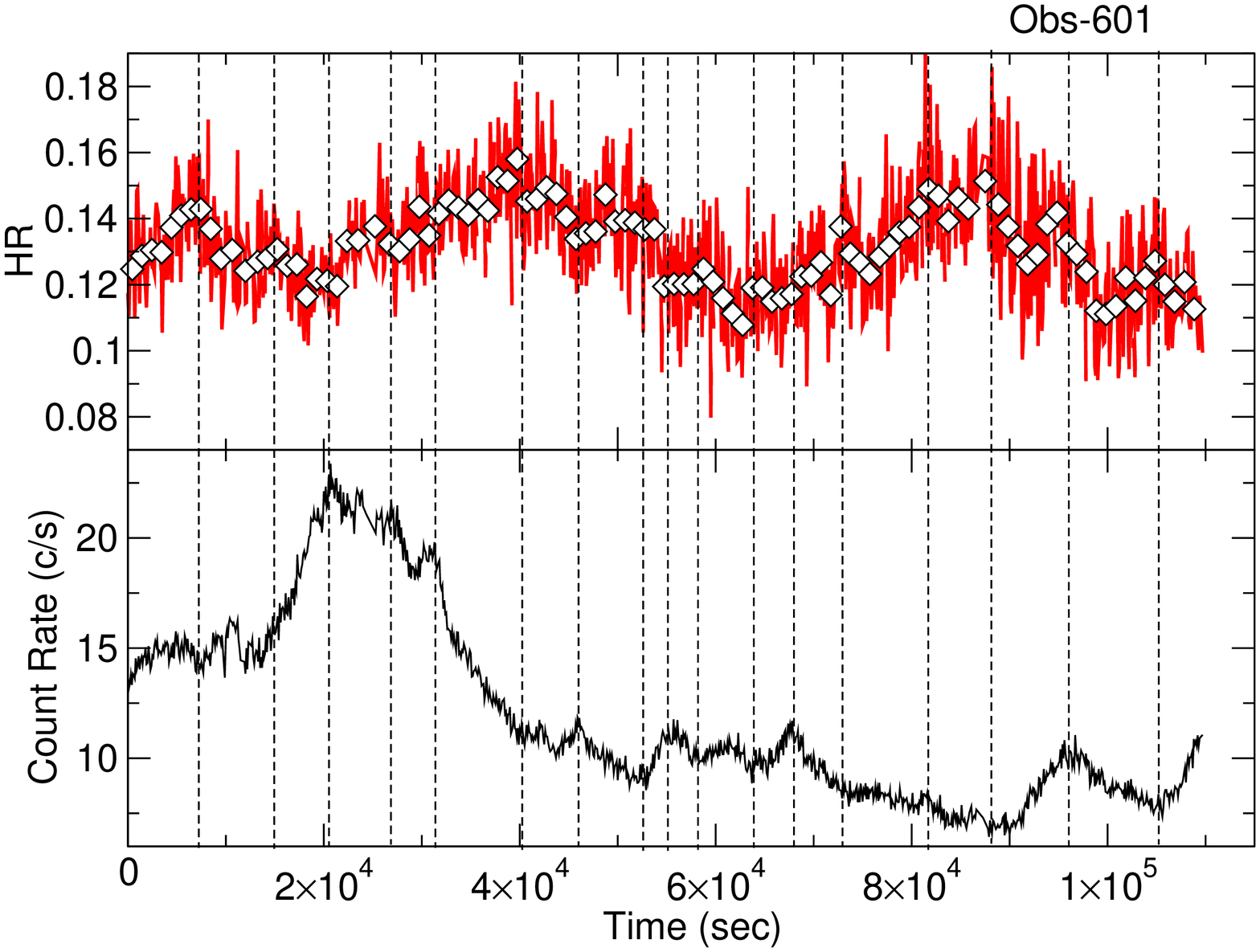,width=8.5cm,height=7cm,%
bbllx=18pt,bblly=35pt,bburx=704pt,bbury=550pt,clip=}
\caption{Top panel: HR plotted as a function of time (obs-601). Points shown in
grey indicate the HR using bins of the size of 100 sec, while open diamonds
indicate the HR values using bins of the size of 1000 sec (for clarity reasons
errors in the latter case are not plotted; they are of the order of $\sim
5\times 10^{-3}$). Lower panel: The full band  light curve (using bins of the
size of 100 sec). The vertical dashed lines indicate the observation  segments
we used for the time resolved spectroscopy (see text in Sect.~5 for details.)}
\label{figure:obs601hr} 
\end{figure}

\section{Spectral variability}
 
The discussion in the above sections suggests that unaccounted detector calibration
inaccuracies may be responsible for the $\sim 0.5-0.6$ keV ``dip" in the residual
plots shown in Fig.~\ref{figure:residua}. However, such inaccuracies most probably
cannot explain the remaining residuals, such as the  the ``wavy" pattern we see in
the two lower panels of the same figure for example. The absence of warm absorbing
signatures cannot explain them either.

Before reaching conclusions regarding the quality of the model fits to the
average spectra of each observation, we have to consider the  fact that, as
shown in Fig.~\ref{figure:tot-lc}, the source exhibits continuous, large
amplitude flux variations on both short and long time scales. If these flux
variations are associated with spectral variations as well (as it is usually 
the case in Seyfert galaxies), then model fitting of some kind of ``average"
spectrum may yield dubious results. 

To illustrate the presence of spectral variations during the \xmm\ observations,
we estimated the ratio of the 2--10 keV over the 0.4--1 keV count rates (i.e.
the ``hardness ratio", or simply HR), using light curves binned in 100 and 1000
sec. The top panel in  Fig.~\ref{figure:obs601hr} shows the HR values for the
obs-601. The respective light curve is plotted in the lower panel. The first
result from this plot is that the HR does not stay constant, i.e., the source
exhibits significant spectral variations, although of an amplitude  smaller than
that of the flux variations: the max-to-min ratio for the HR values is $\sim
1.4$, as opposed to $\sim 3.3$ for that of the observed count rates.

There is also an indication that the spectral variations are anti-correlated with
the flux variations. For example, approximately $7-20$ ksec after the start of
obs-601 the flux increases while HR decreases, and in the next 20 ksec HR increased
while the total source flux decreased. This ``HR--flux" anti-correlation is not
``perfect" though.  For example, around  95--105 ksec after the start of the
observation, HR is  positively correlated with the source flux: they both decrease
with time. A similar behaviour (i.e. significant HR variations, which are mainly
anti-correlated with flux) is seen in the other observations as well (a detailed
study of the spectral variability of the source with the use of HR plots, together
with a cross-correlation analysis, will be presented in a forthcoming paper). 

\section{Time resolved spectroscopy}

To study the spectral variability of the source, we split up obs-201, 301, 401,
501 and 601 in  11, 12, 20, 17 and 18 (78 in total) individual data stretches.
This was done in a way that the segments would broadly correspond to either a
flux-rising or flux-decaying phase. The vertical lines in
Fig.~\ref{figure:obs601hr} indicate the individual segments that we considered
in the case of obs-601. The shortest and longest exposures of a segment in all
observations are 2 ksec and 20 ksec, respectively. As the count rate of the
source is sufficiently large, each of the individual spectra had more than
several ten thousand photons in the 0.4--10 keV band. This is sufficient for an 
accurate determination of the parameters of the relatively simple spectral model
we used to fit them.  Our objectives were to use a simple model to  correctly
parametrise the shape of each one of these spectra, and determine the best-fit
parameter values to investigate whether (and how) the hard and soft band spectra
vary in flux {\it and} shape. 

Regarding the hard band (i.e. 2--10 keV) spectra, a PL model fitted them well.
However, it does not provide an acceptable fit to the 0.4--2 keV band as well. A
soft excess emission component was almost always present at energies below $\sim
2$ keV. We therefore  used the PLBrems model to fit the full band spectra. The
model fit was first performed in the 2--10 keV band. We then kept the  PL slope
and normalisation fixed at their best-fit values when we fit the full band
spectrum. 

\begin{figure}
\psfig{figure=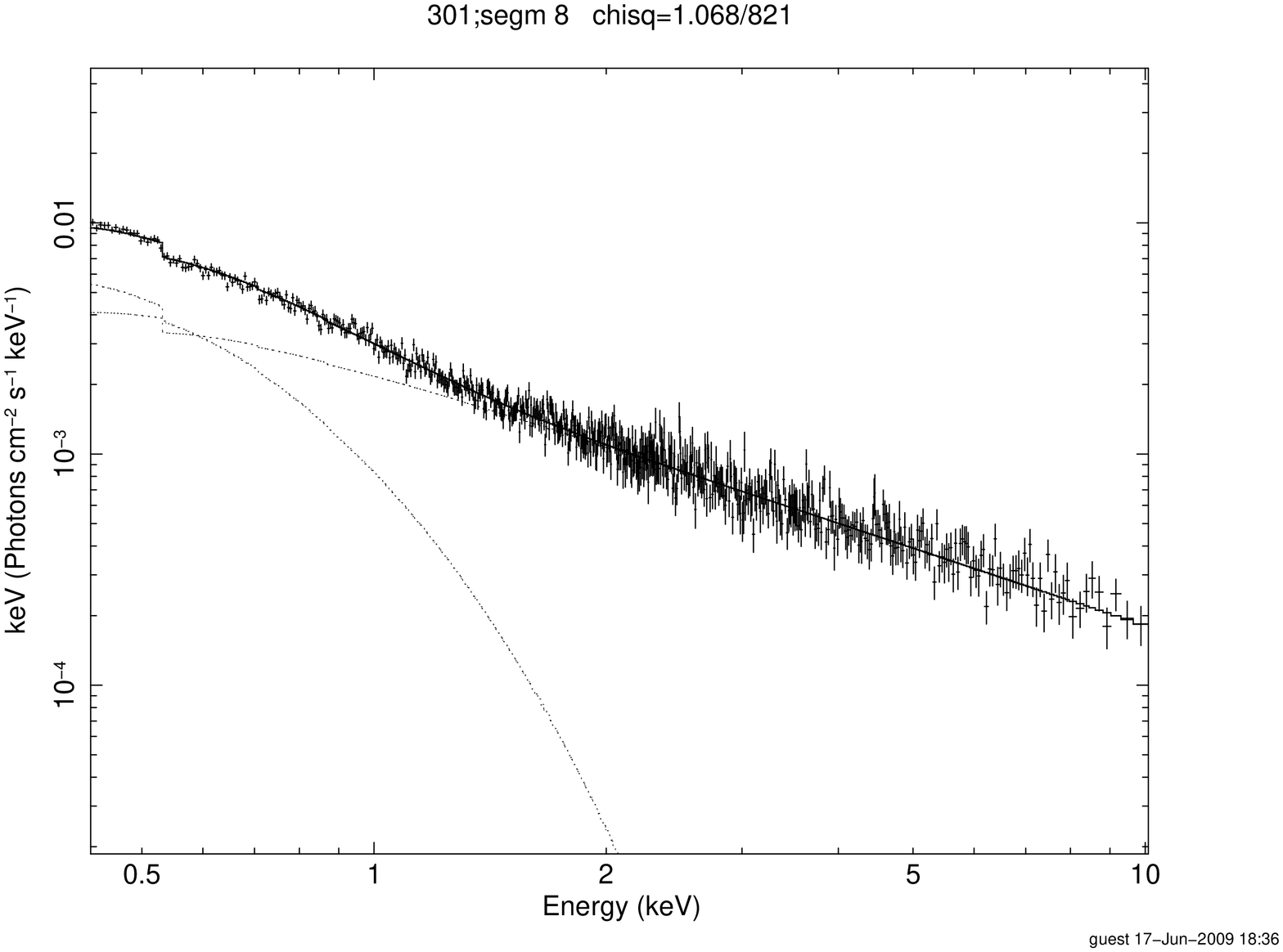,height=5.3truecm,width=8.3truecm,angle=-90,%
 bbllx=525pt,bblly=13pt,bburx=60pt,bbury=708pt,clip=}
\psfig{figure=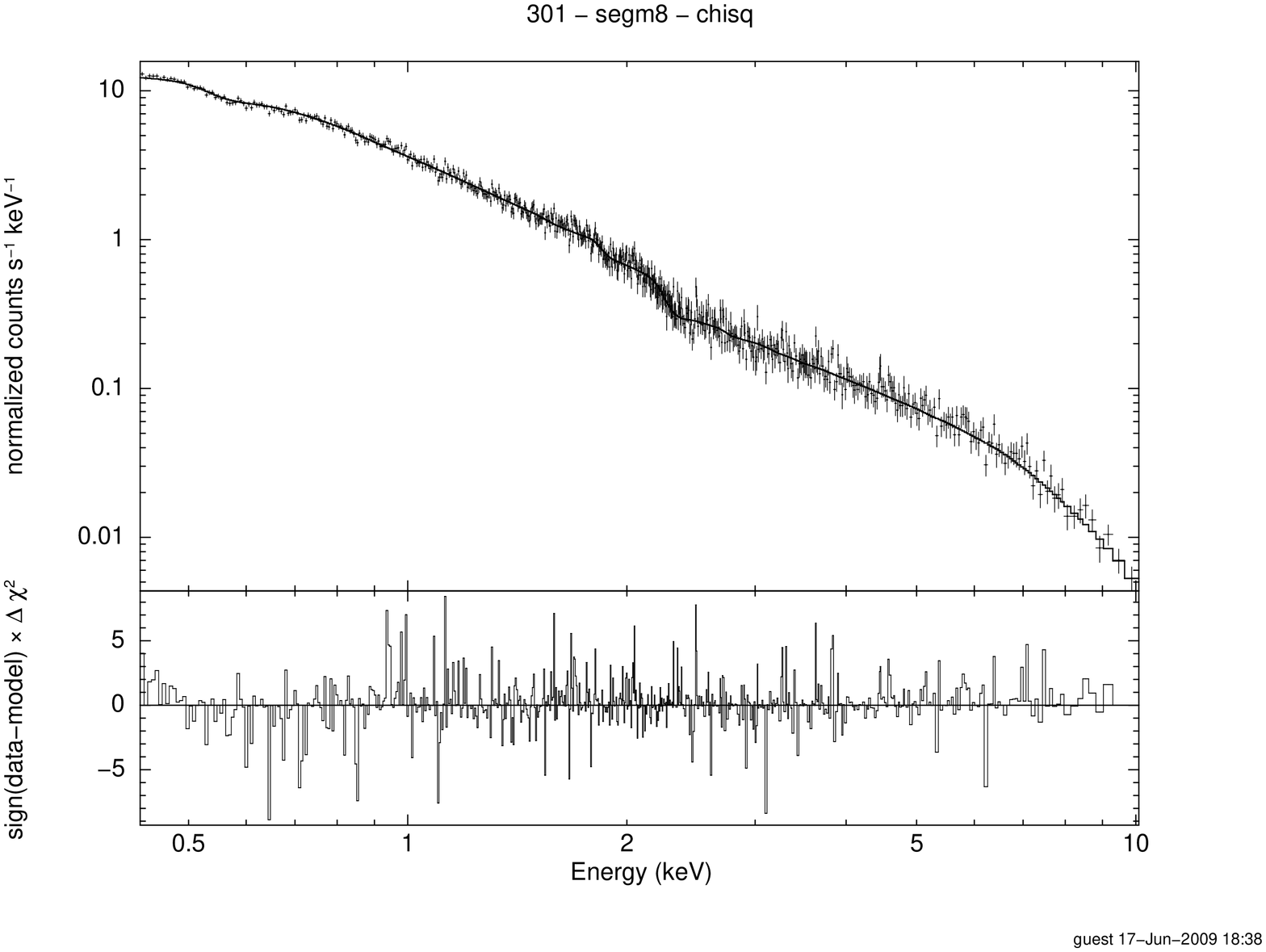,height=2.5truecm,width=8.4truecm,angle=-90,%
 bbllx=550pt,bblly=51pt,bburx=371pt,bbury=720pt,clip=}
\caption[]{Plot of one of the worst PLBrems fits (top panel), and of the
respective residuals (bottom panel) to the PN data of one of the individual
spectra (see Section 5 for details).}
\label{figure:bremsfit}
\end{figure}

The quality of this model fit was statistically acceptable in all cases: the
mean reduced $\chi^2$ of all 78 fitted spectra is $<\chi^2_{\rm red}>$ = 0.991,
with a standard deviation of $\sigma = 0.006$. Figure~\ref{figure:bremsfit}
shows one of the worst PLBrems fits ($\chi^2_{\rm red}=1.068/821$ dof); the data
shown correspond to the time interval between $\sim 51 - 70$ ksec after the
start of obs-301. The dotted lines in the top panel of this figure indicate the
PL component ($\Gamma \sim 2.12$) and the bremsstrahlung emission of $\sim 0.16$
keV which best fitted the spectrum. In the bottom panel we show the contribution
to the total $\chi^2$ per spectral bin.  The residuals plot does not indicate
any systematic large amplitude deviations between the data and the best-fit
model. We reached a similar conclusion when we investigated the residual plots
for all the other spectra as well. 

Although bremsstrahlung emission cannot account for the soft excess emission in
\pks\ (if that were the case we would expect to detect a plethora of emission
lines in the RGS spectrum of the source) our results indicate that this model
can adequately parametrise the shape of this component. We therefore decided to
use the PLBrems best-fit results to study the spectral variability properties of
the source. In particular, we used the best PL spectral slope, $\Gamma$, as a
representation of the hard band spectral shape, and the (un-absorbed) 2--10 keV
and 0.4--1 keV PL flux (Flux$_{\rm PL/2-10  {\rm keV}}$ and  Flux$_{\rm
PL/0.4-1  {\rm keV}}$) as a measure  of the power law continuum flux in the hard
and soft bands, respectively. We also used the  (un-absorbed) 0.4--1 keV flux of
the bremsstrahlung spectral component (Flux$_{\rm Brems}$) as a representative
of the soft excess flux in this object (i.e. of the flux of the excess emission
on top of the hard band PL component, which is assumed to extend down to $\sim
0.4$ keV with the same slope), and the Brems temperature, $k$T, as a 
representative of the soft excess shape. 

\begin{figure}
\psfig{figure=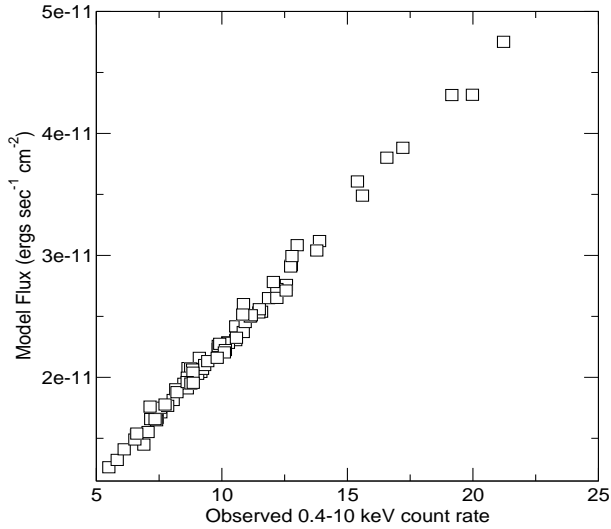,width=8cm,height=7.5cm,%
bbllx=15pt,bblly=248pt,bburx=516pt,bbury=761pt,clip=}
\caption[]{The ``total" model flux (i.e. Flux$_{\rm PL/2-10  {\rm
keV}}+$Flux$_{\rm PL/0.4-1 {\rm keV}}$+Flux$_{\rm Brems}$) plotted as a function
of the observed 0.4--10 keV count rate. }
\label{figure:fluxcntrate}
\end{figure}

As a simple ``sanity" check that our model-fitting results agree with the
observed variations, Fig.~\ref{figure:fluxcntrate} shows the ``total" model flux
(i.e. the sum of Flux$_{\rm PL/2-10 {\rm keV}}$, Flux$_{\rm PL/0.4-1 {\rm keV}}$
and Flux$_{\rm Brems}$) plotted as a function of the mean, 0.4--10 keV count
rate of each segment. The expected one-to-one relation between these two
quantities is obvious from this plot.

%%%%%%%%%%%%%%%%%%%%%%%%%%%%%%%% FIG 9 %%%%%%%%%%%%%%%%%%%%%%%%%%%%%%%%%

%%%%%%%%%%%%%%%%%%%%%%%%%%%%%%%% FIG 10 %%%%%%%%%%%%%%%%%%%%%%%%%%%%%%%%%

\subsection{The hard band spectral variability results} 

\begin{figure}
\psfig{figure=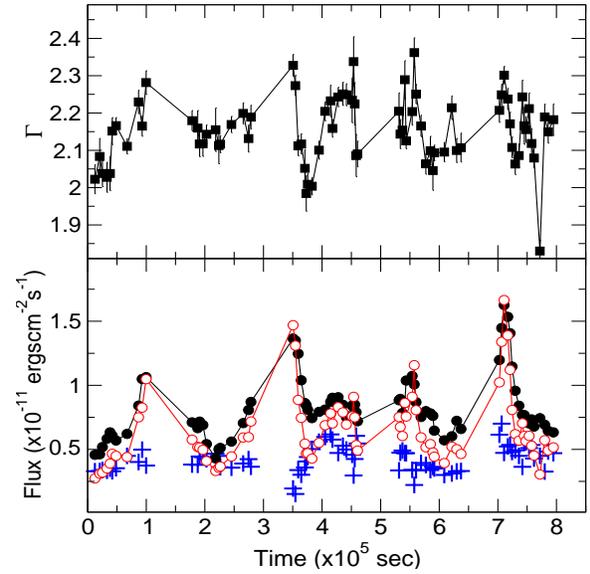,width=8cm,height=7.5cm,%
bbllx=30pt,bblly=248pt,bburx=535pt,bbury=756pt,clip=}
\caption{Top panel: The best-fit spectral slope of the individual spectra of the
source, plotted as a function of time (measured from the start of the \xmm\
observations). Lower panel: Filled, open, and crosses indicate Flux$_{\rm
PL/2-10  {\rm keV}}$, Flux$_{\rm PL/0.4-1  {\rm keV}}$, and Flux$_{\rm Brems}$,
respectively.}
\label{figure:res1}
\end{figure}

\begin{figure}
\psfig{figure=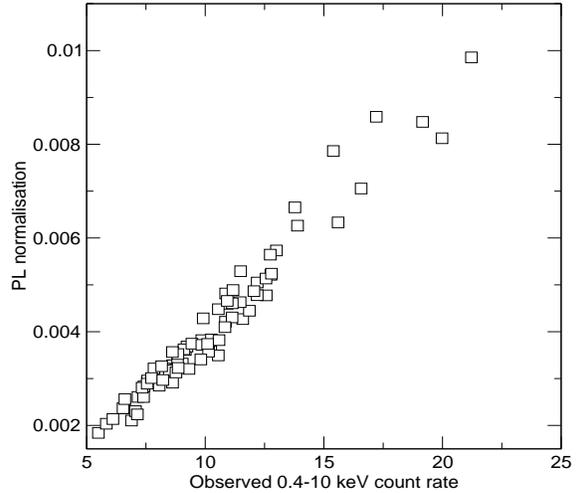,width=7.5cm,height=6.5cm,%
bbllx=20pt,bblly=287pt,bburx=517pt,bbury=756pt,clip=}
\caption[]{The best PL model fit normalisation values plotted as a function of
the  observed 0.4--10 keV count rate. }
\label{figure:plnormcntrate}
\end{figure}

The top panel in Fig.~\ref{figure:res1} displays $\Gamma$ as a function of time
measured from the start of the \xmm\ observations. The power law spectral slope
is significantly variable; we find that $\chi^2=593/77$ dof, when we fit a
constant to the ``$\Gamma$ vs time"  plot.  The average spectral slope is
$\overline{\Gamma}=2.152\pm0.001$. The individual spectral slopes vary between
$\sim 2-2.3$, which implies a max-to-min variability amplitude of $15\%$. The
fractional root mean square variability amplitude (corrected for the
experimental noise) of the spectral slope variations is  $f_{\rm
rms,\Gamma}=3.8\pm 0.1\%$\footnote{The error, accounting only for the
measurement error of $\Gamma$, has been calculated according to the prescription
of Vaughan \etal\ 2003.}. The small amplitude $\Gamma$ variations indicate that
the observed count rate viability is mainly caused by variations of the PL
normalisation. Figure ~\ref{figure:plnormcntrate} shows that this is indeed the
case. 

\begin{figure}
\psfig{figure=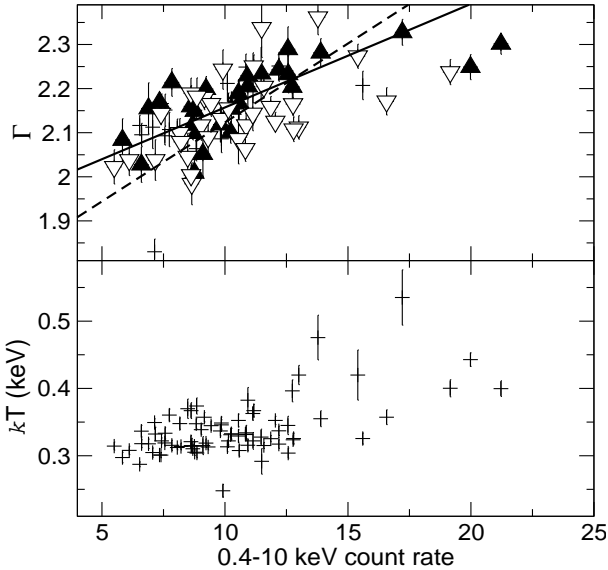,width=8cm,height=7.5cm,%
bbllx=30pt,bblly=285pt,bburx=517pt,bbury=756pt,clip=}
\caption{Top panel: The best-fit spectral slope of the individual spectra of the
source, plotted as the function of the 0.4--10 keV count rate. Filled
``up-triangles" and open ``down-triangles" indicate the $\Gamma_{\rm up}$ and
$\Gamma_{\rm down}$ values, while the solid and dashed lines are the best line
fits to these data sets.  Lower panel: The best-fit Brems $k$T values for the
individual spectra of the source, plotted as the function of the 0.4--10 keV
count rate.}
\label{figure:res2}
\end{figure}

Filled and open circles in the lower panel of Fig.~\ref{figure:res1} indicate
Flux$_{\rm PL/2-10  {\rm keV}}$ and  Flux$_{\rm PL/0.4-1  {\rm keV}}$. A
comparison between the top and bottom panels shows that the spectral slope and
flux variations are highly correlated: the spectrum becomes steeper with
increasing flux. This correlation became clearer when we plotted $\Gamma$ as a
function of the observed $0.4-10$ keV count rate (top panel in
Fig.~\ref{figure:res2}). This is the first time that such a positive
spectral-flux variability correlation (which is commonly detected in other
nearby Seyferts) is  detected for this source. 

%%%%%%%%%%%%%%%%%%%%%%%%%%%%%%%% FIG 11 %%%%%%%%%%%%%%%%%%%%%%%%%%%%%%%%%

%%%%%%%%%%%%%%%%%%%%%%%%%%%%%%%% FIG 12 %%%%%%%%%%%%%%%%%%%%%%%%%%%%%%%%%

%%%%%%%%%%%%%%%%%%%%%%%%%%%%%%%% FIG 13 %%%%%%%%%%%%%%%%%%%%%%%%%%%%%%%%%

Filled (open) triangles in the top  panel of Fig.~\ref{figure:res2} indicate the
spectral slope during the periods when the source flux increases (decreases),
$\Gamma_{\rm up}$ ($\Gamma_{\rm down}$). On average,  the source spectrum
appears to be steeper when the flux increases:  $\overline{\Gamma}_{\rm
up}=2.18\pm0.015$ and $\overline{\Gamma}_{\rm down}=2.14\pm0.015$. But this
difference is not highly significant ($\Delta\overline{\Gamma}=0.04\pm 0.02)$.
The solid and dashed lines in the same panel indicate the best line fits to the
($\Gamma_{\rm up}$, CR) and the ($\Gamma_{\rm down}$, CR) data sets,
respectively. The best-fit lines also imply that, on average, most of the time 
$\Gamma_{\rm up}$ is steeper than $\Gamma_{\rm down}$, but the difference of the
best-fit line parameters is not statistically significant. We therefore cannot
rule out that the shape of the X--ray continuum varies with flux in a similar
way, irrespective of whether the source flux increases or decreases.

\subsection{ The soft excess spectral variability results} Points plotted with
crosses in the lower panel of Fig.~\ref{figure:res1} indicate Flux$_{\rm
Brems}$, plotted as a function of time. The average soft excess flux is $4.23\pm
0.01\times 10^{-12}$\flux, which is smaller than  the average PL flux in the
same band ($\overline{\rm Flux}_{\rm PL/04-1 keV}=6.6\times 10^{-12}$\flux).
Furthermore, the soft excess flux is significantly variable ($\chi^2=47361/77$
dof), but its average amplitude is smaller ($f_{\rm rms,Brems}=23.7\pm0.2$\%)
than the average PL  variability amplitude in the 0.4--1 keV band ($f_{\rm
rms,PL/0.4-1 {\rm keV}}=42.7\pm0.2\%)$. 

The shape of the soft excess is also variable. The best-fit $k$T values are not
consistent with the hypothesis of a constant temperature ($\chi^2=1385/77$ dof).
These variations are of a rather  small amplitude ($f_{\rm
rms,kT}=11.9\pm0.4\%$). Crosses in the bottom panel of Fig.~\ref{figure:res2}
indicate the best-fit $k$T values plotted as function of CR. The soft excess
shape appears to be broadly correlated with the source flux: the temperature
increases with increasing flux. Application of the Kendall's $\tau$ test
resulted in a value of $\tau=0.33$, which  implies a rather weak, but
significant, correlation between $k$T and the source flux (the probability of
the null hypothesis is less than $3\times 10^{-5}$). We conclude that as the
source flux increases $k$T increases as well, i.e. the soft excess ``shifts" to
higher energies. The average $k$T is $0.34\pm 0.01$ keV, which remains the same
irrespective of whether the flux is increasing or decreasing.  The best line
fits to the  ($k$T$_{\rm up}$, CR) and the ($k$T$_{\rm down}$, CR) data sets are
also almost identical. We therefore conclude that the shape of the soft excess
varies with flux in a similar way, irrespective of whether the source flux
increases or decreases. 

\subsection{Cross-correlation between soft excess and hard power law flux} The
bottom panel in Fig.~\ref{figure:res1} indicates that the  PL and Brems flux
light curves are not highly correlated. To investigate this issue further, we
split up all five observations in segments of 2000 sec long (278 in total) and
fitted their 0.4--10 keV band spectra with the PLBrems model.  As above,  the
2--10 keV band was first fitted with a PL model, and we kept the best-fit
$\Gamma$ and PL normalisation values fixed when we added the Brems component to
fit the full band spectra. The model fitted all spectra well, and using the 
best-fit model parameter values we calculated Flux$_{\rm PL/2-10  {\rm keV}}$,
Flux$_{\rm PL/0.4-1  {\rm keV}}$, and  Flux$_{\rm Brems}$. 

\begin{figure}
\psfig{figure=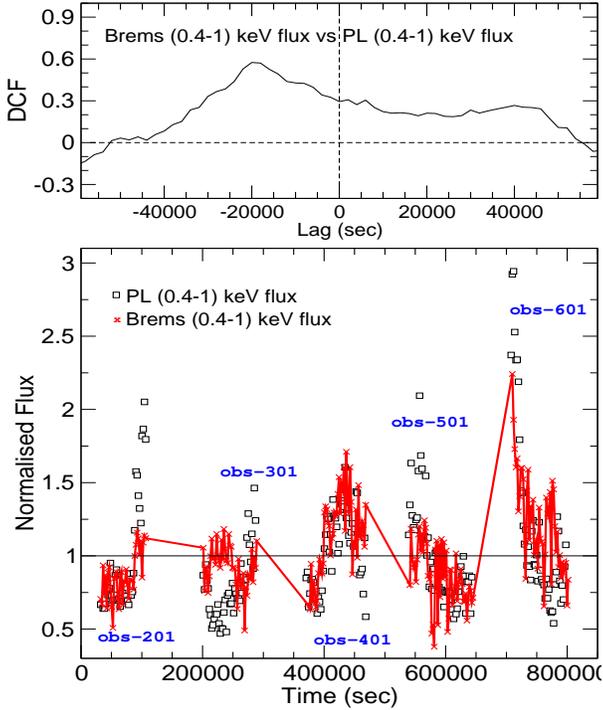,width=8cm,height=9.5cm,%
bbllx=23pt,bblly=100pt,bburx=510pt,bbury=761pt,clip=}
\caption{Top panel: The DCF between Flux$_{\rm PL/0.4-1 {\rm keV}}$ and 
Flux$_{\rm Brems}$. Bottom panel: The respective light curves (open squares and
crosses, respectively), normalized to their mean. The soft excess light curve
has been shifted by +20 ksec, and we plot only the light curve parts which
overlap.}
\label{figure:ccf}
\end{figure}

We determined the cross-correlation between the PL and Brems flux light curves 
using the ``Discrete Correlation Function" (DCF) method of Edelson \& Krolik
(1988). The DCF was calculated using lags of size 2000 sec and in a way that a
significant correlation at positive lags will indicate that  the Flux$_{\rm
PL}$  variations are leading those in the Flux$_{\rm Brems}$ light curve. The
results are plotted in the top panel of Fig.~\ref{figure:ccf}, in the case of
the DCF between the  Flux$_{\rm PL/0.4-1  {\rm keV}}$ and Flux$_{\rm Brems}$.
The maximum DCF amplitude reaches a value of DCF$_{\rm max}\sim 0.6$ at a lag of
lag$_{\rm max}\sim -20$ ksec (the results are  similar when we cross-correlated 
Flux$_{\rm PL/2-10  {\rm keV}}$ and Flux$_{\rm Brems}$, except that  DCF$_{\rm
max}\sim 0.5$ in this case). 

The DCF$_{\rm max}$ value of $\sim 0.6$ implies that the hard band continuum 
and the soft excess flux variations are weakly correlated.  The bottom panel in
Fig.~\ref{figure:ccf} shows  the Flux$_{\rm PL/0.4-1  {\rm keV}}$ and 
Flux$_{\rm Brems}$ light curves (open squares and grey continuous line,
respectively), normalised to their mean.  The soft excess light curve is shifted
by +20 ksec, and we have plotted  only these parts of the light curves which
overlap (i.e., those light curve  parts which correspond to the estimation DCF
at lag$=-20$ ksec). This plot clarifies why the correlation between the hard
band continuum and the soft excess variations is just moderate. The observed
variations agree reasonably well during obs-401, 501 and 601. But the soft and
hard band fluxes are not correlated during obs-301, while the situation is
unclear for obs-201. Furthermore, the difference in the variability amplitude
(even in those cases when the correlation is reasonably good) contributes to the
decrease in the DCF$_{\rm max}$ amplitude.  In any case, even if the two light
curves are indeed intrinsically correlated, the negative time lag implies that
it is the soft excess flux variations which {\it lead} the hard band continuum
variations.

%%%%%%%%%%%%%%%%%%%%%%%%%%%%%%%% FIG 14 %%%%%%%%%%%%%%%%%%%%%%%%%%%

\section{The full band PN spectrum revisited}

Given the complex spectral variability of the source, we chose seven of the 78
segments into which we originally split the total \xmm\ observation, based on
the following criteria: i) HR remained constant during the exposure of each
segment, ii) the average count rate is different and adequately samples the
observed flux variability amplitude of the source, and iii) there are enough
photons in the spectra to ensure a sufficiently high signal to noise ratio. The
first and second columns in Table~\ref{table:fitfin} list the start time, the
exposure time, and the average count rate of the segments we chose (segments are
listed in da ecreasing order of the average count rate). There are more than
$\sim 60,000$ photons in the 0.4--10 keV band of each one of these spectra.

First, we fitted the 2--10 keV spectrum of each segment with a PL model
(best-fit slopes, $\Gamma_{\rm hard}$, are listed in the third column of
Table~\ref{table:fitfin}), which fitted the high energy part of the  spectra
well. Then we kept $\Gamma_{\rm hard}$ and the best-fit PL normalisation fixed,
and we fitted the 0.4--10 keV spectra with the following more physically
motivated additional models:

%%%%%%%%%%%%%%%%%%%%%%%%%%%%%%%% FIG 15 %%%%%%%%%%%%%%%%%%%%%%%%%%%%%
\begin{figure}
\psfig{figure=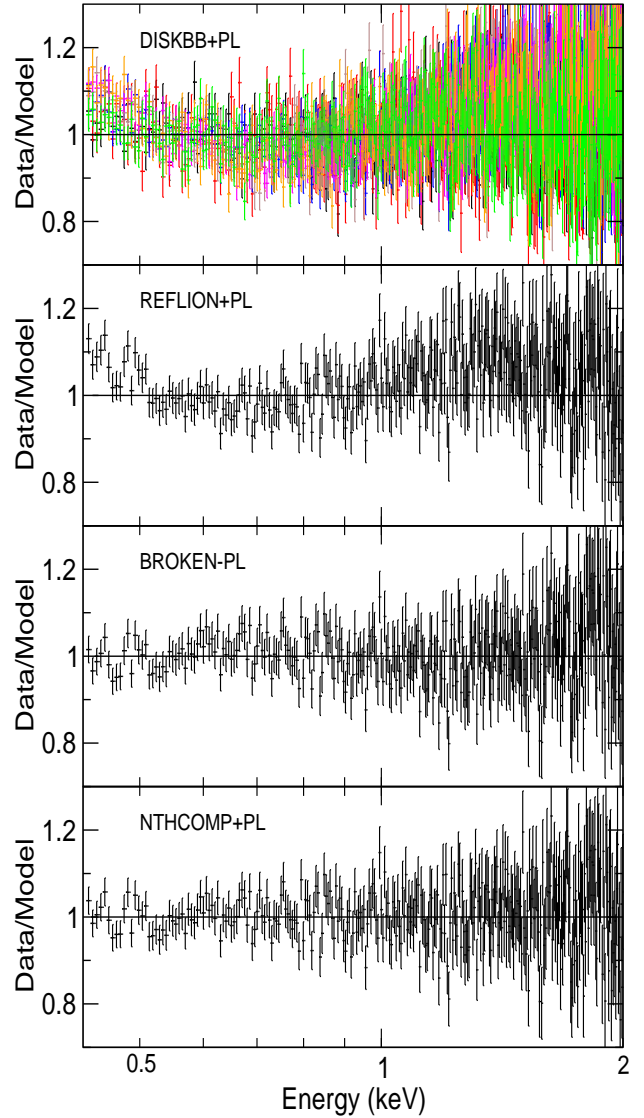,height=15truecm,width=8.5truecm,%
 bbllx=25pt,bblly=40pt,bburx=520pt,bbury=760pt,clip=}
\caption{From top to bottom: Data-to-model ratio in the 0.4$-$2 keV band  for
the best fits of a {\tt DISKBB}+PL, {\tt REFLION}+PL, a broken PL, and  a {\tt
NTHCOMP}+PL model to the seven individual spectra we considered in Sect. 6. The
residuals for the spectra listed in Table 3 are plotted with points in black,
red, green, blue, brown, magenta and orange, respectively. It is difficult to
identify the individual residual plots for each spectrum, because they are
similar for all of them. For clarity reasons, in the following panels  we plot
the best-fit residuals in the case of the spectrum of the fourth segment listed
in Table~\ref{table:fitfin} only. The residuals are similar for the other
spectra as well. }
\label{figure:finres}
\end{figure}

a) a standard accretion disc, multi black body spectrum, i.e. {\tt DISKBB}. The
best-fit temperature at the inner disc radius, T$_{\rm in}$, turned out to be
similar in all cases. We therefore re-fitted the spectra simultaneously by
keeping T$_{\rm in}$ the same for all seven spectra. The model gave a poor fit
to the data ($\chi^2=6027/5122$ dof). The best-fit residuals are plotted in the
top panel of Fig.~\ref{figure:finres}. They are plotted  with points in black,
red, green, blue, brown, magenta and orange for the spectra as listed in
Table~3.  They are very similar for all spectra, and for that reason it is
difficult to spot the residuals for each spectrum in this panel.

b) a relativistically blurred photo-ionised reflection model. We used the  model
of Ross \& Fabian (2005) (which is available as an additional table model,  {\tt
REFLION}, in {\tt XSPEC}), in which a semi-infinite slab of optically thick cold
gas of a constant density is illuminated by a PL continuum.  Although we did not
detect any significant evidence for disc  reflection  in the hard band spectrum
of the source, we considered this model, because strong blurring effects from
the relativistic accretion disc can smear the iron line almost completely and at
the same time result in a featureless soft excess emission. For each spectrum, 
we kept the PL continuum slope fixed at  the $\Gamma_{\rm hard}$ values listed
in Table~\ref{table:fitfin}, and we  added the illuminating and reflected
components (with the amount of each free to vary). To simulate the blurring from
the disc, we convolved the total emission with a Laor (1991) line profile ({\tt
KDBLUR} model in {\tt XSPEC}). We fitted the seven spectra simultaneously, with
the same iron abundance of the disc (relative to solar). We also fixed the inner
radius of the Laor profile to be 1.235 gravitational radii (the smallest radius
allowed in the case of a maximally rotating BH), and we also assumed a common 
emissivity profile and the  disc's inclination angle. The best-fit values were:
emissivity index  $\sim 10$ (which is the  maximum value allowed in {\tt
KDBLUR}), inclination $\sim 72$ degrees, and iron abundance $\sim 1.8$. The
ionization parameter of the gas was decreasing from $\xi\sim 3750$ to $\xi\sim
500$ for the highest to the  lowest  flux spectrum, respectively. The final 
best model fit was very poor ($\chi^2=7316.4/5119$ dof). In the second panel of
Fig.~\ref{figure:finres}, we plot the best-fit residuals in the case of the
spectrum of the fourth segment listed in Table~\ref{table:fitfin} only. The
residuals are similar for the other spectra as well, and for clarity reasons we
do not plot them.

c) a BKNPL model. This model is frequently used to parametrise the X--ray
spectra of BL Lac objects (see e.g. Brinkmann \etal\ 2005). The spectra were
fitted simultaneously by keeping the hard band slope fixed at $\Gamma_{\rm
hard}$.  The best-fit ``break energy", E$_{\rm break}$,  turned out to be $\sim
2$ keV in all cases, except for the highest flux spectrum, where E$_{\rm
break}\sim 2.9$ keV. The best-fit slope at soft energies was decreasing form
$\sim 2.7$ to $\sim 2.5$, from the low to the high flux spectra. The best model
fit is  statistically acceptable in this case ($\chi^2=5081/5115$ dof). The data
plotted in the third panel of Fig.~\ref{figure:finres} indicate the  best-fit
residuals in the case of the  spectrum of the fourth segment listed in
Table~\ref{table:fitfin}. The residuals have a similar shape and amplitude in
the other spectra as well.

d) a low-temperature Comptonisation model. We first fitted the spectra
individually, using the {\tt compTT} model. The best-fit T$_{\rm 0}$ and  $\tau$
values turned out to be quite similar for all spectra. For that reason, we
re-fitted the spectra simultaneously, assuming the same T$_{\rm 0}$ and  $\tau$.
The resulting best-fit electron temperature values were increasing from $kT\sim
2.8$ keV, in the case of the largest flux spectrum,  to $\sim 5.4$ keV, for the
lowest flux spectrum. The fit was statistically acceptable, with a $\chi^2$
value of 5124 for 5120 dof. However, the best-fit residuals plot was
qualitatively similar to the {\tt compTT}+PL model residuals plotted in
Fig.~\ref{figure:residua}: they had a similar ``wavy" pattern, with a deficit
around 0.5--0.7 keV, although of significantly smaller amplitude. 

To further investigate  the issue of low-temperature Comptonisation, we also
considered the model {\tt NTHCOMP} in {\tt XSPEC} (Zycki, Done \& Smith 1999),
which also describes the continuum shape from thermal Comptonisation. The
parameters of the model are: a) the seed photon temperature, $k$T$_{\rm bb}$, 
b) the electron temperature, $k$T$_{\rm e}$, and c) the asymptotic power law
photon index, $\Gamma_{\rm Comp}$ (which is set by the combination of the
electron scattering optical depth and $k$T$_{\rm e}$). We first fitted the seven
spectra separately (assuming a disc-black body soft photon input). We noticed
that the individual best-fit $k$T$_{\rm bb}$ and  $k$T$_{\rm e}$ values were
similar. We therefore re-fitted the spectra simultaneously,  assuming the same
$k$T$_{\rm bb}$ and  $k$T$_{\rm e}$. The model fit was very good, with a
$\chi^2$ of 4997 for 5120 dof  (the best-fit residuals plot shown in the bottom
panel of Fig.~\ref{figure:finres} is for the spectrum of the fourth segment
listed in Table~\ref{table:fitfin}; similar  residuals were observed for the
other spectra as well). The best-fit electron and seed photon temperatures were
$k$T$_{\rm e}=0.424\pm0.012$ keV and $k$T$_{\rm bb}=0.11^{+0.004}_{-0.01}$ keV.
The best-fit $\Gamma_{\rm Comp}$ values for this model are listed in
Table~\ref{table:fitfin}. 

\begin{table}[]
\small
\tabcolsep1ex
\caption{\label{table:fitfin} Results from the spectral fits of the {\tt
NTHCOMP}+PL  model to the 0.4--10 keV spectra of the seven short segments (see
section 6 for details), assuming the same electron temperature ($k$T=0.42$\pm
0.01$ keV), and disc-black body  seed photon temperature ($k$T$_{\rm
bb}=0.11^{+0.004}_{-0.01}$ keV), in all cases. Columns 1 and 2 list details of
the segments we used.}
\begin{tabular}{lccc}
\noalign{\smallskip} \hline \noalign{\smallskip}
\multicolumn{1}{l}{Time} &  \multicolumn{1}{c}{Exp.~/~$\overline{\rm CR}$} & 
\multicolumn{1}{c}{$\Gamma_{\rm hard}$} &
\multicolumn{1}{c}{$\Gamma_{\rm Comp}$}  \\
\multicolumn{1}{c}{day/UT} & \multicolumn{1}{c}{ksec~/~(c/sec) } & 
\multicolumn{1}{c}{}&
\multicolumn{1}{c}{} \\
\noalign{\smallskip} \hline \noalign{\smallskip}
2008-09-15; 05:33:50 & 4.2/19.98 & 2.26$\pm0.05$ & $2.68\pm0.07$ \\
2008-09-15; 04:30:23 & 3.8/15.60 & 2.21$\pm0.05$ & $3.31\pm0.11$ \\
2008-09-08; 02:59:53 & 6.8/12.77 & 2.21$\pm0.04$ & $3.65\pm0.15$ \\
2008-09-11; 21:22:40 & 9.8/11.86 & 2.10$\pm0.04$ & $3.10\pm0.06$ \\
2008-09-11; 11:39:12 & 12/8.62   & 1.95$\pm0.04$ & $3.22\pm0.06$ \\
2008-09-09; 21:28:49 & 20/7.34   & 2.05$\pm0.04$ & $3.08\pm0.05$ \\
2008-09-14; 04:22:24 & 12.9/6.60 & 2.01$\pm0.04$ & $3.05\pm0.06$ \\
\noalign{\smallskip}\hline
\end{tabular}
\medskip
\end{table}

Both the BKNPL and the {\tt NTHCOMP}+PL models gave a good fit to the data.
However, in the case of the  latter model, we got a reduction in the best-fit
$\chi^2$ value of $\Delta\chi^2=84$ with a {\it smaller} number of model
parameters. Following Mushotzky (1982), we defined the ratio of the likelihood of
the best BKNPL fit, $L_{\rm BKNPL}$,  to the likelihood of the best {\tt
NTHCOMP}+PL model fit, $L_{\rm Comp}$, as:  $L_{\rm BKNPL}/L_{\rm
Comp}=$exp[($\chi^2_{\rm BKNPL}-\chi^2_{\rm Compt})/2$]=exp$(\Delta\chi^2/2)\sim
2\times 10^{18}$, in our case. This high value strongly suggests that the
Comptonisation model is much more likely than the BKNPL model to be the
``correct" model for the soft excess in \pks.  

\section{Summary and discussion}

We presented the results from the spectral analysis of a long \xmm\ observation
of the luminous, radio-loud, narrow line Seyfert 1 galaxy \pks. The object was
observed in September 2008 for 5 orbits, yielding data of unprecedented quality
for this source. The main results from our work are summarised below:

1) The source exhibits continuous, large amplitude flux variations on all
sampled time scales. A flux variability study with the use of the power
spectrum analysis will be presented in a companion paper. 

2) The 2--10 keV band spectrum of the source is well fitted by a simple PL
model, with a steep slope of $\Gamma\sim 2.15$, which is typical of NLS1.  

3) The hard band slope is variable; the spectrum steepens as the flux increases.
This is similar to what is commonly observed in other Seyferts as well (see e.g.
Sobolewska \& Papadakis, 2009). 

4) The hard band spectrum is very smooth; we did not observe any absorption
features (either lines or edges), and we found only weak evidence for the
presence of an iron line, which is indicative  of  emission from highly ionised
iron. The line's EW is small, $\sim 20$ eV, with an 90\% upper limit of no more
than $\sim 45-50$ eV.

5) At energies below $\sim 2$ keV a strong, steep, smooth, and  broad soft
excess component appears. The soft excess flux is variable, but with an
amplitude which is smaller than the hard band continuum variability amplitude.
The soft excess spectra shape is also variable and becomes ``harder" with
increasing flux. 

6) The soft excess and the hard band flux variations are moderately correlated,
but with a delay: the former are leading the latter by $\sim 20$ ksec.

7) We detected no signs of  warm absorption features at energies below $\sim 2$
keV.

8) The average 0.4--10 keV spectra of the individual \xmm, orbit-long
observations  cannot be fitted well by any of the models that are commonly used
to fit similar data of other NLS1. This could be partly due to remaining
uncertainties in the instrumental PN calibration.  But we believe that the
spectral variability of the source, and in particular the variable soft/hard
flux ratio (even on time scales of the order of a few ksec), significantly
affects the the results from the spectral analysis of spectra which are averaged
on time scales longer than $10-20$ ksec. 

9) Model-fitting to the spectrum of a few short segments of the total
observations showed that the soft excess can be well described by a
low-temperature Comptonisation model of the disc photons, in agreement with the
results of Brinkmann \etal\ (2004). Other alternatives like emission from a
multicolour disc black body, or reflection from an ionised disc, could not fit 
the data well. A broken power law model also provided a statistically acceptable
fit to the data, however, the low-temperature Comptonisation model is a more
likely explanation of the source spectrum.

Although \pks\ is one of the few radio-loud NLS1, its X--ray spectrum seems to
be ``normal" and rather similar to  the spectra of other NLS1. Our results rule
out a jet origin for the bulk of the X--ray emission from this object: the hard
band slope of $\sim 2.15$ is similar to the X--ray slope of most radio-quiet
NLS1s, the ``soften when brighter" spectral variability trend of the power law
continuum is identical to what is observed in other radio-quiet Seyferts, and
even the weakness of the iron line emission is also consistent with what is
observed in other NLS1s. Furthermore, the fact that we found  no evidence for
ionised absorption in either the soft or the hard  X-ray bands implies that
the  X-ray spectrum of \pks\ is essentially unmodified by intervening matter. As
a result, the present data set is ideal to study the intrinsic, X--ray emission
spectrum of this object, free from absorption effects. Given the similarity
between the X--ray emission properties between \pks\ and other NLS1s, some of
the results from the present study may have wider implications to NLS1 as a
class. 

However, \pks\ may still be a rather rare object, because it almost certainly
has a high mass accretion rate, $\dot{\rm m}$. To estimate this rate in
Eddington units, $\dot{\rm m_{\rm Edd}}$, we need to know the mass of the BH in
the nucleus of \pks. We can estimate it  by using the virial relation M$_{\rm
BH}=R(\Delta V)^2/G$, where $R$ and $\Delta V$ are the radius and velocity
dispersion of the broad line region. We measured the velocity dispersion from
the width of the broad H$\beta$ emission line (Corbin, 1997) and the radius of
the broad line region from the   radius--luminosity scaling relation of Bentz et
al. (2009), using the  5100 \AA\ luminosity measurement of Corbin \& Smith
(2000), and we found that M$_{\rm BH}\sim 5.8\times 10^7$ M$_{\odot}$.  The
average 2--10 keV flux of $\sim 1.8\times 10^{-11}$ \flux\ (Gliozzi \etal\ 2007)
implies an average 2--10 keV luminosity of $\sim 8.4\times 10^{44}$ ergs/s,
which is already $\sim 11\%$ of the Eddington luminosity (L$_{\rm Edd}$) for
this BH. Even if we use the lowest bolometric correction of $\sim 20$ of
Vasudevan \& Fabian (2009), the bolometric accretion rate of the source turns
out to be larger than the Eddington accretion rate, $\dot{\rm m}_{\rm Edd}$. In
fact, the combined analysis of the present data set with simultaneous SWIFT
observations implies that the source is accreting at a rate which may be several
times larger than $\dot{\rm m}_{\rm Edd}$ (Gliozzi et al. in preparation).
Keeping this fact in mind, we present below a discussion of some of the results
we reported above.

{\it The hard band X--ray spectral variability of \pks.}  Despite the high
accretion rate of the source, its hard band spectrum is not unusual. It has a
power law shape, which, as we mentioned in Sect. 3.1, can be explained by
thermal Comptonisation of seed photons by a hot electron corona whose parameters
are similar to those which can also explain the hard band X-ray spectrum of most
radio-quiet Seyferts. 

In the case of thermal Comptonisation, the spectral slope $\Gamma$ is 
determined by the so-called Compton amplification factor, $A=(L_{\rm
diss}+L_{\rm s})/L_{\rm s}$, where  $L_{\rm diss}$ is the power dissipated in
the corona and $L_{\rm s}$ is the intercepted soft luminosity. Obviously, if $A$
varies, then $\Gamma$ should also change as well. According to Beloborodov
(1999), $\Gamma \propto (A-1)^{-b}$, where $b\approx 0.1$, if the energy of the
input soft photons, E$_{\rm in}$,  is of the order of a few eVs, and $b\approx
0.17$, if E$_{\rm in}$ is of the order of a few hundred eV. 

The data plotted in the top panel of Fig.~\ref{figure:res1} can be well fitted
by a power law model of the form: $\Gamma\propto$ X--ray$_{\rm flux}^{0.15\pm
0.02}$ (here  X--ray$_{\rm flux}$ stands for the observed 0.4--10 keV count
rate). A possible explanation then for the ``spectral slope - X--ray flux"
correlation is that the ``heating-to-cooling" ratio of the corona, $(L_{\rm
diss}/L_{\rm s})$, decreases with increasing X--ray flux (or accretion rate, if
the X--ray flux is indicative of the accretion rate of the source), i.e. 
$(L_{\rm diss}/L_{\rm s})\propto$X--ray$_{\rm flux}^{-1}$. In this case, we
would expect that  $\Gamma \propto$ X--ray$_{\rm flux}^{b}$, with $b\approx
0.17$ in the case when the  energy of the input photons is of the order of a
hundred eV or so, exactly as observed. 

The hard band spectral variations in \pks\ could also be  explained by a
combination of a highly variable (in flux) power law continuum  (with a fixed
slope) and a constant reflection component (e.g. Taylor \etal\ 2003; Ponti
\etal\ 2006; Miniutti \etal\ 2007), or by variations in the column density,
covering fraction, and/or the opacity of an absorber, while the continuum slope
remains constant (e.g. Miller, Turner, \& Reeves, 2008; Turner \etal\ 2007).
Still, our results indicate that the X--ray spectrum of \pks\ is not modified by
intervening absorbing material, while the failure of the reflection model to
account well for the full band spectrum of the source (in various flux levels)
implies that the reflection component is probably minimal in the 2--10 keV band
(when compared to the PL continuum). For these reasons, we believe that the
observed spectral variations in this source indicate {\it intrinsic} $\Gamma$
variations.  The fact that  the ``steeper when brighter" variability pattern of
\pks\ is similar to what is observed in many Seyfert galaxies suggests that
intrinsic $\Gamma$ variations may also be responsible to a large extend for the
spectral variations we observe in these objects as well.

{\it The iron emission line.} We observe some weak evidence for the presence of
a line with an EW $\sim 20$ eV at $6.8\pm 0.1$ keV, consistent with the
K$\alpha$ line from He-like iron. \pks\ is not unusual among radio-quiet objects
in this respect, as similar lines have been observed in many AGN. Bianchi \etal\
(2009) for example report the detection of such a line with an EW less than 20
eV in at least 12 (broad and narrow line) objects in their sample. The weakness
of the line's strength may be easily explained if the disc is covered by a hot
layer which is up-scattering the disc photons to produce the observed
soft-excess (see below). If its optical depth is larger than unity, then even if
the disc is producing iron lines due to reflection, they will be significantly
suppressed by Compton scattering in this layer (Matt \etal\ 1997). 

We do not observe any evidence from iron line emission of neutral material. The
99\% upper EW limit for a narrow iron K$\alpha$ line in the total spectrum of
the source is just 11 eV. This is  significantly smaller than the EW of $\sim
40$ eV that is expected from objects of luminosity $\sim 8\times 10^{44}$ ergs/s
(Bianchi \etal\ 2007). In the case of lines arising from reflection of X--rays
from the disc, this result can be explained from the fact that  the disc
material is highly ionised, as discussed above. But iron lines can also be
produced by reprocessing of X--rays from distant, cold Compton-thick matter, as
for instance from the putative torus. The lack of a neutral iron K$\alpha$ line
suggests that the covering  factor of the Compton-thick torus in this object
must be significantly smaller than the  covering factor in similar luminosity
Seyferts. On the other hand, the absence of this line agrees with the fact that
\pks\ is most probably accreting at a rate higher than $\dot{\rm m}_{\rm Edd}$.
Indeed, objects with $\dot{\rm m}$ significantly higher than $\dot{\rm m}_{\rm
Edd}$ show lines with an EW which is smaller than 20-10 eV (Bianchi \etal\
2007). Perhaps the increase of the opening angle of the torus may then depend
mainly on the accretion rate and not just the luminosity of the source.

{\it The soft excess.}  We found that the soft excess emission below $\sim 2$
keV is best explained by a low-temperature ($\sim 0.42$ keV) Comptonisation
component, where the temperature of the seed photons is $\sim 0.11$ keV. This 
agrees with the results of Brinkmann \etal\ (2004). The seed photons temperature
is not consistent with the expected maximum  temperature in the innermost region
of a Shakura \& Sunyaev, optically thick and geometrically thin accretion disc
around a  $\sim 6\times 10^7$ M$_{\odot}$ BH (which should be $\sim 15$ times
smaller than $\sim 0.1$ keV in the case when $\dot{\rm m}\sim 0.3\dot{\rm
m}_{\rm Edd}$). However, in the case of super-Eddington accretion discs, the
accretion flow deviates from the thermodynamic equilibrium, and as a result the
accreted gas can be overheated in the central region. Beloborodov (1998) has
shown that the maximum temperature in the disc around a massive (Schwarzschild)
black hole can be as large as $\sim 10^7$ K, if the accretion rate is 100 times
$\dot{\rm m}_{\rm Edd}$. Since the temperature of the gas in the disc determines
the emission spectrum to a large extent, it may not be surprising for the
spectrum of a disc with a super-Eddington accretion rate to extend up to 0.1 keV
or so, as observed. 

Furthermore, Kawaguchi (2003) has demonstrated that a hot layer (where
Comptonisation becomes important)  can be developped in the inner region of
super-Eddington discs when $\dot{\rm m}>10\dot{\rm m}_{\rm Edd}$. Perhaps then,
due to its super-Eddington accretion rate, there is a hot layer on top of the
disc in \pks, which can be identified as the  low-temperature Comptonising
medium responsible for its soft excess emission. Interestingly, low temperature
Comptonisation spectra can also fit well the soft excess emission in two AGN
which also have a super-Eddington rate, namely  EX J0136.9-3510 and RE J1034+396
(Jon \etal\ 2009; Middleton \etal\ 2009). If this is the case,  then perhaps the
physical origin of the soft excess in these objects may be different from the
origin of the same component in NLS1 with a sub-Eddington accretion rate,
because we would not expect the presence of hot layers in these objects where 
significant Comptonisation can distort the underlying disc spectrum.

We also found that both the soft excess flux and shape are variable. Regarding
the spectral variability of this component, we found that  its  spectral shape 
``hardens" (i.e. it shifts to higher energies) with increasing flux (lower panel
in Fig. 11). According to the model-fit results presented in the previous
section, this spectral ``hardening" can be explained by a decrease (i.e.
flattening) of $\Gamma_{\rm Comp}$ at a constant electron temperature with
increasing flux.  Since $\Gamma_{\rm Comp}$ depends on the optical depth of the
Comptonising medium as well,  the $\Gamma_{\rm Comp}$ flattening implies an
increase in $\tau$ with increasing flux. Similar spectral variations were also
reported by Brinkmann \etal\ (2004), who  also attributed them to $\tau$
variation. 

In physical terms, an increase in the normalisation of the soft excess  can be
achieved by an increase in the size of the emitting area. Indeed, the radial
extent of the hot disc layer, where Comptonisation effects become important,
does increases with increasing $\dot{\rm m}$ (i.e. increasing flux) according
to  Kawaguchi (2003). At the same time the Compton $y$ parameter (which depends
on the temperature and the optical depth due to scattering in the hot layer) 
increases as well, and since $\Gamma \propto y^{-2/9}$ (Beloborodov, 1999),
$\Gamma_{\rm Comp}$ should flatten with increasing $\dot{\rm m}$, as observed.
Consequently, both the shape of the soft excess component and its flux and
spectral variations are consistent with the hypothesis that, due to the
super-Eddington accretion rate of the source, a hot disc layer exists in the
innermost region of \pks, which significantly distorts the underlying disc
emission. An increase in the accretion rate will result in an increase of the
size of the hot layer as well as hence an increase in the flux of the soft
excess, and an increase in the Compton $y$ parameter, which will in turn result
in the ``hardening" of the shape of this component. 

{\it The delay between soft and hard band flux variations.} Our results suggest
that the soft excess flux may lead the hard band X--ray flux variations by $\sim
20$ ksec. If real, such a correlation cannot be explained by reflection models;
therefore this result is another indication that such models cannot account for
the soft excess component in \pks. 

We argued in the paragraph above that the soft excess flux variability may be 
caused by $\dot{\rm m}$ perturbations. One possible explanation for the
correlation and the time delay between soft and hard band variations is that 
the hard band X-rays are produced at a radius $r_{\rm hard}$, which is smaller
than the radius where most of the soft excess emission originates, $r_{\rm
soft}$, and the $\dot{\rm m}$ perturbations can propagate from $r_{\rm soft}$ to
$r_{\rm hard}$. The time-delay between the soft excess and hard band X--rays
could then be identified with the propagation/diffusion time scale of the
perturbations, which should be of the order of $t_{\rm diff}\sim r_{\rm
soft}/v_{r}$, where $v_{r}$ is the radial drift velocity of the accretion flow.
For example, according to Kawaguchi (2003), in a $\dot{\rm m}=100\dot{\rm
m}_{\rm Edd}$ source, the hot layer, which could be  responsible for the soft
excess emission, should extend from $r\sim 100$ to $r\sim 2$ (in units of the
Schwarzschild radius). If most of the 0.4--1 keV soft excess photons originate
at $r\sim 10$, then $t_{\rm diff}\sim 10-30$ ksec. If furthermore  the hard
X--rays are mainly produced at $r\le 3$, then the delay we detected could be
identified with this $t_{\rm diff}$ value. 

We note though that even if this is indeed the case in \pks, these $\dot{\rm m}$
perturbations should affect the soft and hard band Comptonising regions in a
different way; the soft excess ``hardens"  while the hard band spectrum
``softens" with increasing flux (i.e. $\dot{\rm m}$). If the $\dot{\rm m}$
perturbations mainly affect the optical depth of the scattering  media, they do
it in such a way that while $\tau$ increases in the hot disc layer, it should
decrease in the hot corona which is responsible for the hard band X--ray
continuum. Clearly, more work is needed to elaborate further on the relation
between the soft and hard band spectral and flux variations in \pks, and other
Seyferts as well. 

\begin{acknowledgements} 

IE and WPB acknowledge partial support from the EU ToK grant 39965 and  
FP7-REGPOT 206469. M.G. acknowledges support by the XMM-Newton Guest
Investigator Program under NASA grant 201593. This work is based on observations
with XMM-Newton, an ESA science mission with instruments and contributions
directly funded by ESA Member States and the USA (NASA). This research has made
use of the NASA/IPAC Extragalactic Database (NED) which is operated by the Jet
Propulsion Laboratory, California Institute of Technology, under contract with
the National Aeronautics and Space Administration.

\end{acknowledgements}

\end{document}